%% file: root.tex
\documentclass[letterpaper, 10 pt, conference]{ieeeconf}
\IEEEoverridecommandlockouts
\usepackage{cite}
\usepackage{amsmath,amssymb,amsfonts}
\usepackage{mathrsfs}
\usepackage{algorithmic}
\usepackage{graphicx}
\usepackage{textcomp}
\usepackage{xcolor}
\let\labelindent\relax
\usepackage{enumitem}

\allowdisplaybreaks

\usepackage{amsthm}

\usepackage{times}
\usepackage{bm}
\usepackage{mathtools}

\usepackage{url}
\usepackage[colorlinks=true, linkcolor=blue, citecolor=blue, urlcolor=blue]{hyperref}
\usepackage{cleveref}  

\usepackage[justification=raggedright, singlelinecheck=false]{caption}

\newif\iftogglecontent
\togglecontenttrue  % Set to \togglecontentfalse to disable content

\newcommand{\toggle}[1]{\iftogglecontent #1\fi}
\newcommand{\toggletext}[2]{%
    \iftogglecontent
        #1%
    \else 
        #2%
    \fi
}

\def\BibTeX{{\rm B\kern-.05em{\sc i\kern-.025em b}\kern-.08em
    T\kern-.1667em\lower.7ex\hbox{E}\kern-.125emX}}

\newtheorem{prp}{Theorem}
\crefname{prp}{Theorem}{Theorems}
\newtheorem{lmm}{Lemma}
\crefname{lmm}{Lemma}{Lemmas}
\newtheorem{df}{Definition}
\crefname{df}{Definition}{Definitions}

\crefname{as}{Assumption}{Assumptions}
\newtheorem{rem}{Remark}
\crefname{rem}{Remark}{Remark}

\newcommand{\R}{\mathbb{R}}
\newcommand{\T}{\mathsf{T}}
\DeclareMathOperator*{\argmin}{arg\,min}

\begin{document}

%\title{Certified Control of Latent Dynamics\\
\title{Latent Representations for Control Design with \\ Provable 
Stability and Safety Guarantees\\

\author{Paul Lutkus, Kaiyuan Wang, Lars Lindemann, and Stephen Tu 
% \thanks{Paul Lutkus, Kaiyuan Wang, Lars Lindemann, and Stephen Tu are with the Department of XXXX. 
%         {Email: \tt\small XXX} }
% }
\thanks{P. Lutkus is with the Thomas Lord Department of Computer Science, University of Southern California. L. Lindemann is with the Automatic Control Laboratory, ETH Zürich. K. Wang and S. Tu are with the Ming Hsieh Department of Electrical and Computer Engineering, University of Southern California. Corresponding E-mail: \url{lutkus@usc.edu}}
\thanks{This work was partially supported by the National Science Foundation through NSF CPS \#2434460.}
}
}

\maketitle

\begin{abstract}
% We consider the problem of learning latent representations of dynamical systems for verifiable control synthesis. Specifically, our goal is to map state observations into lower-dimensional latent spaces to compute control inputs that guarantee stability and safety. Computing control inputs via low-dimensional latent representations can allow for the application of control and verification tools, such as Lyapunov or barrier functions, that might otherwise be computationally expensive when applied to the full state representation. We start by learning latent  dynamics, encoders, and decoders and provide two conditions, referred to as conjugacy conditions, over the  system dynamics and these learned functions. Together with a control certificate constructed in the latent space (e.g., a barrier or a Lyapunov function), these conjugacy conditions guarantee the desired system behavior (e.g., safety or stability). One of our insights is that enforcing these conjugacy conditions during training can greatly improve control performance. We provide two case studies: (1) the stabilization of a cartpole system, and (2) collision avoidance of a two-car system. To the best of our knowledge, this is the first work to learn latent representations for control that do provide safety and stability guarantees.
%
We initiate a formal study on the use of low-dimensional latent representations of dynamical systems for verifiable control synthesis.
Our main goal is to enable the application of  verification techniques---such as Lyapunov or barrier functions---that might otherwise be computationally prohibitive when applied directly to the full state representation.
Towards this goal, we first provide dynamics-aware, approximate conjugacy conditions which formalize the notion of reconstruction error necessary for systems analysis. 
We then utilize our conjugacy conditions to \emph{transfer} the stability and invariance guarantees of a latent certificate function (e.g., a Lyapunov or barrier function) for a latent space controller back to the original system.
Importantly, our analysis contains several important implications for learning latent spaces and dynamics, by highlighting the necessary geometric properties which need to be preserved by the latent space, in addition to providing concrete loss functions for dynamics reconstruction that are directly related to control design.
We conclude by demonstrating the applicability of our theory to two case studies: (1) stabilization of a cartpole system, and (2) collision avoidance for a two-vehicle system.
% To the best of our knowledge, this is the first work providing formal guarantees for control design from latent representations.
\end{abstract}

\section{Introduction}
In control theory, we are primarily interested in ensuring stability and safety for dynamical systems---e.g., designing a stabilizing controller for an industrial manipulator or a safe controller for a drone operating in a cluttered environment.
Fortunately, the control community has developed a variety of tools for this purpose, such as
% For instance, a common way of certifying stability and safety is via
%
Lyapunov \cite{freeman2008robust} and barrier functions \cite{ames2014control} for certifying stability and safety, respectively. However, constructing these functions in general is NP-hard. Indeed, we can only construct these functions analytically for limited classes of systems, while computational tools such as sum-of-squares programming \cite{prajna2004safety,prajna2007framework} or learn \& verify paradigms involving satisfiability modulo theory solvers \cite{kapinski2014simulation,abate2020formal} are prohibitive  for complex, higher-dimensional systems due to computational intractability.  
Furthermore, these challenges are present more broadly, e.g., in model predictive control and Hamilton-Jacobi reachability~\cite{mitchell2005time}.

As the design of verifiable controllers becomes specifically challenging when the system dimension increases---e.g., for multi-agent or systems with large relative degree---a natural question to ask towards addressing these scalability challenges is: \emph{can we instead design controllers in a lower-dimensional space, while ensuring that any verifiable guarantees still  apply to the original dynamics?} Classic work in model order reduction provides answers when systems have simple structures, e.g., for linear interconnected systems \cite{obinata2012model,sandberg2009model}, and when
the dynamics are known. When the system is more complex or accurate models are not available, such classic techniques do not generalize.

Motivated by these shortcomings, 
% in combination with the successes of representation learning techniques from the machine learning community, 
% recently 
there has been substantial interest in utilizing representation learning techniques such as auto-encoders~\cite{bengio2013representationlearning} to learn lower dimensional latent representations where Lyapunov and barrier functions can be applied.
However, a crucial missing link in this growing body of literature is how to \emph{transfer} guarantees which arise from analysis of control policies in the latent space to 
corresponding guarantees in the original state space. 
In this work, we study the connections between latent space analysis and original-space guarantees~(\Cref{fig:latent_safe_set}). In particular, we first   
define dynamics-aware approximate conjugacy conditions 
% involving both the encoder/decoder
% and the latent/original dynamics models, 
which serve as proper measures of reconstruction error
that are suitable for systems analysis. 
We then use these conditions to transfer
both asymptotic stability and set-invariance guarantees---\emph{to the extent allowed by the geometry of the latent space}---from a latent-space based controller in feedback with a latent dynamics model, to an induced original-space controller in feedback with the true dynamics.
% by composing the latent-space controller with the encoder. 

\toggletext{
% FULL VERSION VVV
\begin{figure*}
\centering
\includegraphics[width=0.9\columnwidth]{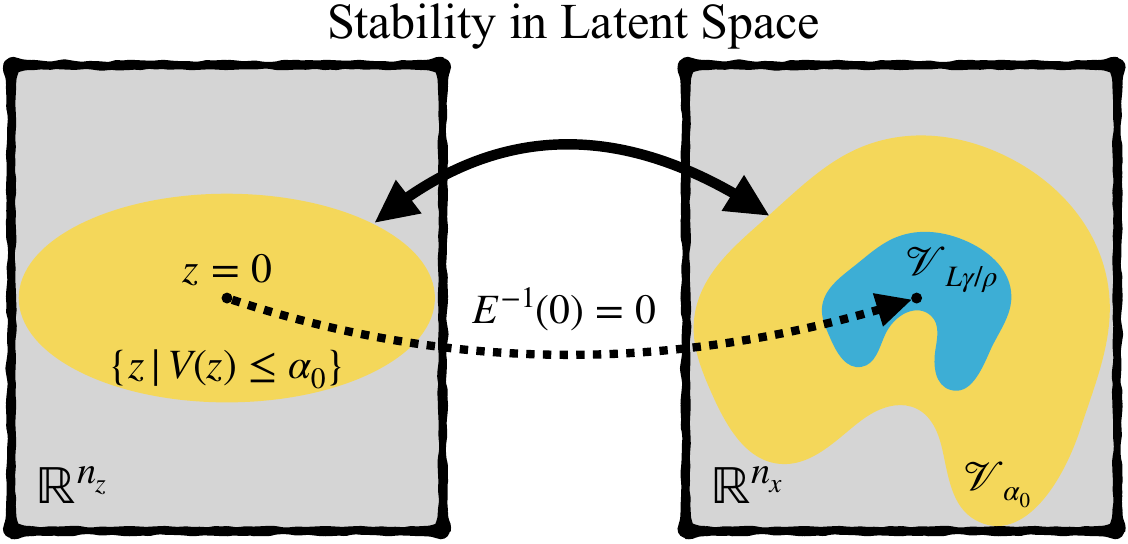}
\includegraphics[width=0.9\columnwidth]{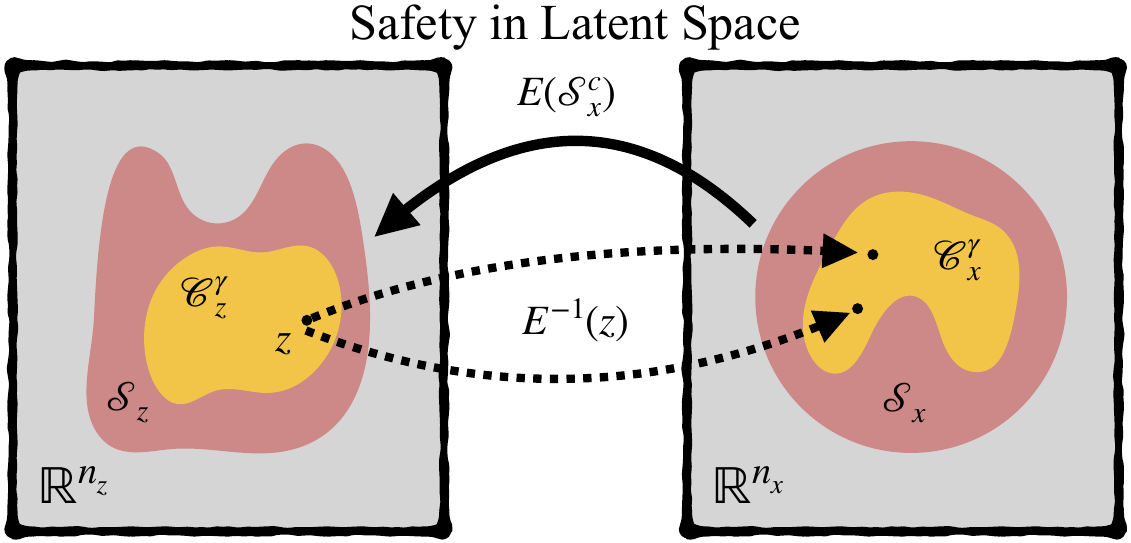}
\caption{\textbf{Left (Stability)}: A Lyapunov function $V$ in latent space $\R^{n_z}$ guarantees: (1) stability of the origin, and (2) forward invariance of sub-level sets of $V$ (shown in yellow). The Lyapunov function $V$ induces in the original space $\mathbb{R}^{n_x}$: (1) forward invariant sets (shown in yellow), and (2) an attractive set $\mathcal{V}_{L\gamma/\rho}$ that contains the origin (shown in blue). \textbf{Right (Safety)}: The set $\mathcal{S}_x$ denotes a safe set in the original space $\R^{n_x}$ that induces a safe set $\mathcal{S}_z$ in the latent space $\R^{n_z}$ (shown in red); we have that $E^{-1}(\mathcal{S}_z)\subseteq \mathcal{S}_x$. A barrier function $h$  in latent space guarantees forward invariance of a set $\mathcal{C}_z^\gamma \subset \mathcal{S}_z$ (shown in yellow). The set $\mathcal{C}_z^\gamma$ induces a forward invariant and safe set $\mathcal{C}_x^\gamma$ in the original space, i.e., $\mathcal{C}_x^\gamma\subseteq \mathcal{S}_x$. 
}
\label{fig:latent_safe_set}
\end{figure*}}{% CDC VERSION VVV
\begin{figure*}
\centering
\includegraphics[width=0.8\columnwidth]{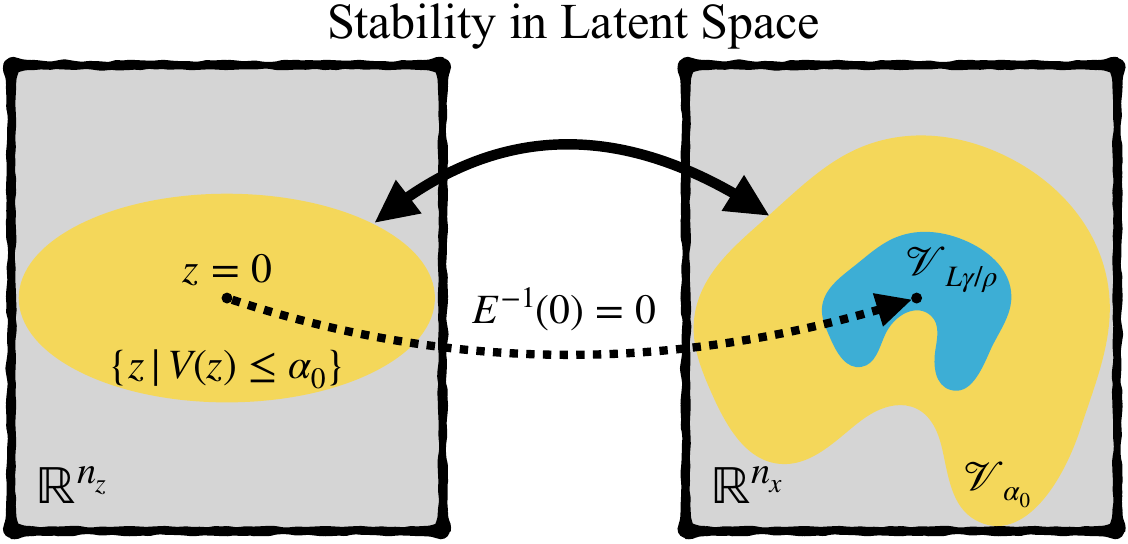}
\includegraphics[width=0.8\columnwidth]{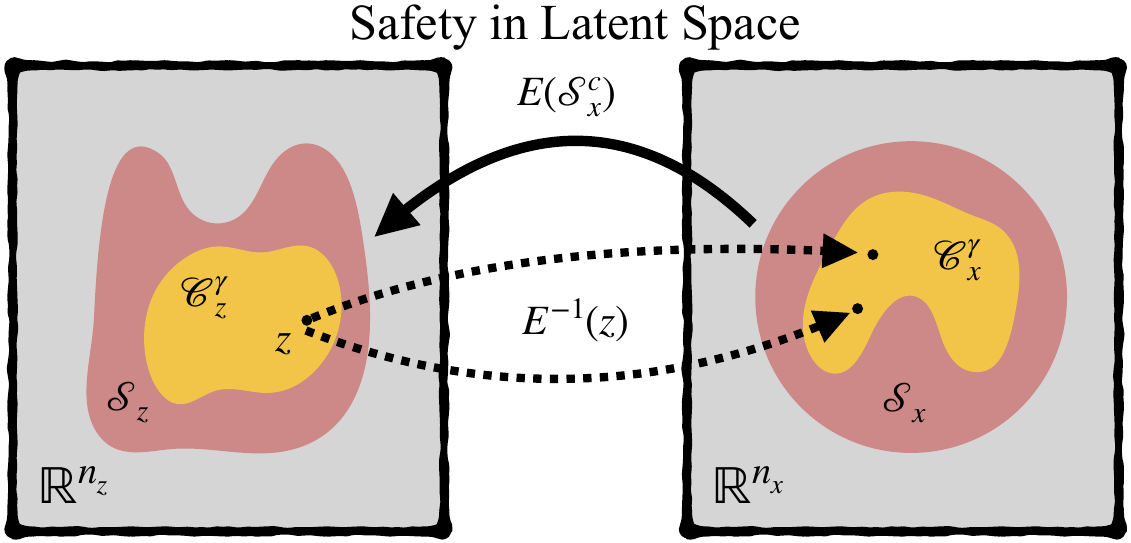}
\caption{\small{\textbf{Left (Stability)}: Latent Lyapunov function $V$ on $\R^{n_z}$ guarantees: (1) stability of the origin, and (2) forward invariance of sub-level sets of $V$ (yellow), and induces in the original space $\mathbb{R}^{n_x}$: (1) forward invariant sets, and (2) an attractive set $\mathcal{V}_{L\gamma/\rho}$ containing the origin. \textbf{Right (Safety)}: The safe set $\mathcal{S}_x\subseteq\R^{n_x}$ induces a safe set $\mathcal{S}_z$ in the latent space $\R^{n_z}$ where $E^{-1}(\mathcal{S}_z)\subseteq \mathcal{S}_x$. Barrier function $h$ on $\R^{n_z}$ guarantees forward invariance of a set $\mathcal{C}_z^\gamma \subset \mathcal{S}_z$, and induces a forward invariant set $\mathcal{C}_x^\gamma$ in the original space, i.e., $\mathcal{C}_x^\gamma\subseteq \mathcal{S}_x$.}
}
\label{fig:latent_safe_set}
\end{figure*}}

Our work has several implications when learning latent spaces and dynamics models for control. First, we make explicit the geometric properties which are required to be preserved by the latent space in order to transfer stability and invariance guarantees back to the original space. 
Second, our conjugacy conditions yield concrete loss functions for dynamics reconstruction which are directly tied to the downstream control task.
%; we show experimentally that 
% using these loss functions indeed benefits downstream control performance.
% }
% \textcolor{red}{
We make the following contributions:
\begin{enumerate}[noitemsep,leftmargin=*,topsep=0pt, parsep=0pt]
    \item We give formal definitions of approximate dynamics conjugacy which capture
    the interaction between the encoder/decoder and the latent/original dynamics.
    \item We characterize the relationship between approximately conjugate dynamics and the satisfaction of Lyapunov and barrier inequalities, allowing us to precisely state when a latent space control design stabilizes or renders a safe-set invariant for the original system. 
    \item We illustrate the application of our theory for both latent Lyapunov and barrier functions
    for stabilizing a cartpole and
    collision avoidance in a two vehicle system.
    %but also 
    % demonstrate the relative advantages of our proposed conjugacy losses over
    % baseline approaches to learning latent spaces and corresponding latent dynamics. 
\end{enumerate}
% In order to provide a general set of results, 
We conduct analysis for both discrete and continuous-time systems; our specific conjugacy definitions differ between the two settings, highlighting the importance of treating both.
% The full paper with omitted proofs is available in~\cite{lutkus2024latent}.

%\begin{figure}[ht]
%\centering
%\includegraphics[width=0.95\columnwidth]{l4dc2025_latex_template/figures/latent_safety_sets.pdf}
%\caption{A visualization of the geometry of the latent safe %set $\mathcal{S}_z$ from \eqref{eq:latent_safe_set}.
%In the right display the original $\R^{n_x}$ space with its safe set $\mathcal{S}_x$ is shown.
%In the left display the induced latent safe set $\mathcal{S}_z$ is shown; we have that $E^{-1}(\mathcal{S}_z)\subseteq \mathcal{S}_x$.
%The set $\mathcal{C}_z^\gamma$
%is the forward invariant set under
%the latent dynamics, and $\mathcal{C}_x^\gamma$ is its induced forward invariant set under the original dynamics.
%}
%\label{fig:latent_safe_set}
%\end{figure}

%\begin{figure}[ht]
%\centering
%\includegraphics[width=0.95\columnwidth]{l4dc2025_latex_template/figures/latent_stability_sets.pdf}
%\caption{A visualization of the sets from %\Cref{continuous_lyapunov_proposition}.
%In the left display the latent $\R^{n_z}$ space with the sub-level set of the Lyapunov function $V$ at level $\alpha_0$  is shown. In the right display the original $\R^{n_z}$ space, the induced forward invariant set $\mathcal{V}_{\alpha_0}$, and the attractive set $\mathcal{V}_{L\gamma/\rho}$ is shown. 
%}
%\label{fig:latent_stability_set}
%\end{figure}

\subsection{Related Work}

\noindent\textit{Learning Lyapunov and barrier functions.} Towards efficient computation of stability and safety certificates, recent work has proposed to learn candidate Lyapunov and barrier functions. However, these learned candidates are not guaranteed to be valid by construction and have to be verified, a process which is computationally expensive.
%
% The complexity here is especially hidden in the expensive verification step. 
For the purpose of verification, various methods have been proposed, e.g., satisfiability modulo theory solvers \cite{kapinski2014simulation,abate2020formal}, symbolic bound propagation methods \cite{hu2024verification}, convex optimization problems when convex candidate functions are used \cite{chen2024verification,chen2021learning}, statistical verification \cite{boffi2020learning,tayal2025cp}, and Lipschitz constants~\cite{montenegro2024data}. 
Importantly, this learn-then-verify paradigm applies in a similar way to finding control Lyapunov and barrier functions, see e.g., \cite{robey2020learning,dawson2022safe}.
While these techniques are typically deployed in the original state space of the system, they can also verify latent space designs, for which our analysis can be used to transfer guarantees back to the original space.

% from which the
% guarantees can transferred over to the original space utilizing our analysis. 

% whenever our proposed conjugacy conditions are satisfied.

\noindent\textit{Latent planning, control, and analysis.}
The use of latent spaces for planning and control is an active area of research in the ML/robotics communities. A major focus is on enabling control from visual observations by encoding images into latent spaces with latent dynamics, for which techniques such as model-predictive control~\cite{watter2015e2c,wilcox2021ls}, 
reinforcement learning~\cite{hafner2019planet}, 
% motion planning~\cite{ichter2019latentrrt}, 
and HJ reachability~\cite{nakamura2025generalizing} can be applied in latent space. While effective empirically, these works do not provide stability and invariance guarantees for the closed-loop system.
In \cite{vieira2024morals}, Morse graph analysis is performed on a learned latent space in order to determine regions of attraction. 
The works \cite{castaneda2023latentcbf,kumar2024latentcbf,tayal2024semi}  propose algorithmic frameworks to learn latent CBFs for safe control, and \cite{tayal2024semi} analyzes when their learning algorithm returns a  CBF for the original system. Our work is a generalization of these results that abstracts away the learning process and focuses on identifying the conjugacy conditions that enable transfer guarantees.

\noindent\textit{Reduced order models (RoM).}
Classic RoM methods focus on linear dynamical systems and their interconnections~\cite{obinata2012model,sandberg2009model,Volkwein2013}.
For non-linear systems, RoMs are typically defined in a domain-specific and case-by-case manner, such as SLIP models~\cite{shahbazi2016slip} for robotic locomotion. A related concept involves the decomposition of high-dimensional dynamics into simplified models 
from the literature on templates (maximally-reduced models) and anchors (the higher-dimensional counterparts that templates model), see e.g., \cite{full1999templates}.
Non-linear RoM models have also been used in conjunction with
safety-critical CBFs~\cite{choen2024cbfrom,molnar2023rom,cohen2024safety}.
Obtaining the RoM or low-dimensional manifold, and its associated tracking controller, is typically done case-by-case. 
Regarding analysis of RoM systems,\cite{astolfi2003immersion} 
describes conditions for asymptotically 
immersing a high-dimensional system onto an invariant low-dimensional manifold, and
\cite{tabuada2008approximate} gives sufficient conditions in terms of incremental stability of sub-systems
for the existence of an approximate RoM with bounded tracking error. The latter work shares similar motivations as ours, and we will further discuss the connections between this work and ours in \Cref{sec:approximate_conjugacy}.

%
% take inspiration from robotic locomotion, where simple models
% such as SLIP models~\cite{shahbazi2016slip} capture enough fidelity of the full non-linear system
% to enable performant control design.
%

Recently, there has been a push for learning-based solutions
to broaden the scope of RoM models. 
Approaches such as 
autoencoders~\cite{masti2021learning},
spectral methods~\cite{alora2023data,alora2023robust}, and
reinforcement learning (RL)~\cite{chen2024reinforcement}
have been used to obtain RoMs from data,
which are then controlled using model-predictive control (MPC). 
Similarly, \cite{compton2024learning} learns a predictive model for the RoM discrepancy, which is used to robustify control.
%
%
% In \cite{chen2024reinforcement}, reduced order models for bipedal walkers are learned via reinforcement learning (RL), and then controlled using a model-predictive controller (MPC). 
% % \toggle{\textcolor{blue}{Autoencoders are used to learn state representations and output maps from input-output data, for MPC, in \cite{masti2021learning}.}} 
% Similarly, both autoencoders~\cite{masti2021learning} and 
% spectral methods~\cite{alora2023data,alora2023robust}
% have also been used to obtain reduced order models from data.
% % which are controlled via MPC controllers. 
Furthermore, \cite{mcinroe2024global} extends the templates and anchors framework into a learning context in order to compute hierarchies of attracting sets from dynamical data.
We also briefly mention Koopman operator techniques~\cite{brunton2022koopman}, which can also be viewed through the lens of RoMs, as they provide spectral approximations for modeling general non-linear systems.
Our work is again complementary, as we define a general set of conditions for which RoM models can be used for control design, and do not prescribe a specific model parameterization nor controller.

\noindent\textbf{Notation.} 
Let $\circ$ denote function composition, i.e., $(f_2\circ f_1)(x):=f_2(f_1(x))$ for two functions $f_1:\mathbb{R}^{n_1}\mapsto\mathbb{R}^{n_2}$ and $f_2:\mathbb{R}^{n_2}\mapsto\mathbb{R}^{n_3}$. The Jacobian of $f:\mathbb{R}^{n_1}\mapsto\mathbb{R}^{n_2}$ evaluated at $x_0 \in \R^{n_1}$ is denoted $\frac{\partial f}{\partial x}(x_0)=\partial f(x_0)$. The Moore-Penrose pseudo-inverse of a matrix $A$ is denoted $A^{\dagger}$.
% with full row-rank to be $A^\dagger=A^T(AA^T)^{-1}$.
The $\ell_2$-norm of a vector $x\in\R^{n_x}$ is denoted by $\|x\|$.

\section{Problem Formulation}
Let $x_t\in\mathbb{R}^{n_x}$ and $u_t\in\mathbb{R}^{n_u}$ be the state and the control input of our dynamical system at time $t$.\footnote{We consider discrete and continuous time, i.e., either $t\in\mathbb{R}_{\ge 0}$ or $t\in\mathbb{N}$.} Specifically, let
\begin{align}\label{eq:system}
    x_t^+= f(x_t, u_t), \quad x_0 \in \R^{n_x},
\end{align}
model the system dynamics described by the function $f:\mathbb{R}^{n_x}\times\mathbb{R}^{n_u}\mapsto\mathbb{R}^{n_x}$. We use $x_t^+ := \dot{x}_t$ to denote the time-derivative of $x_t$ in continuous time and $x_t^+:= x_{t+1}$ to denote the next state in discrete time. Our goal is to design feedback control policies $\bar{\pi} : \R^{n_x} \mapsto \R^{n_u}$ so that the closed-loop system 
$x_t^+ = f(x_t, \bar{\pi}(x_t))$ has stability or safety properties. To design such a control policy $\bar{\pi}$, we consider a particular approach based on \emph{latent-space dynamics}.  

The pipeline we consider works as follows. First, we obtain an encoder/decoder pair $(E, D)$ where $E : \R^{n_x} \mapsto \R^{n_z}$ and $D: \R^{n_z} \mapsto \R^{n_x}$ link the original state space $\R^{n_x}$ to a latent state space $\R^{n_z}$, and vice versa. We consider settings where $n_z \leq n_x$, so that the latent-space serves as a form of \emph{compression} of the original state space; an ideal pair hence satisfies $E \circ D \approx \mathrm{Id}$ for the states relevant for control.  Second, we obtain a latent-space dynamical system $f_z:\mathbb{R}^{n_z}\mapsto \mathbb{R}^{n_z}$ so that $z_t^+ = f_z(z_t, u_t)$ with latent state $z_t=E(x_t) \in \R^{n_z}$. Lastly, we perform control design in the latent space --- involving standard techniques such as (control) Lyapunov and barrier functions --- and obtain a latent space controller $\pi : \R^{n_z} \mapsto \R^{n_u}$. This latent space controller $\pi$ induces a controller $\bar{\pi}$ in the original state space via the composition $\bar{\pi}=\pi \circ E$. We are agnostic to how both the encoder/decoder pair $(E, D)$ and the latent dynamics $f_z$ are obtained; they can either be learned using auto-encoders~\cite{bengio2013representationlearning}
or specified using domain-specific knowledge.

This approach yields two key closed-loop dynamics:
\begin{align}
    z_t^+ &= f_z^{(\pi)}(z_t) := f_z(z_t, \pi(z_t)), \\
    x_t^+ &= f^{(\pi)}(x_t) := f(x_t, (\pi \circ E)(x_t)).
\end{align}
The main motivation for this setup is that, as the latent space dimension satisfies $n_z \leq n_x$, 
from a computational perspective the latent closed-loop dynamics $f_z^{(\pi)}$ are potentially much more amenable to formal verification and numerical analysis techniques, compared with any formal analysis in the original state space dimension $n_x$.
However, a formal guarantee on $f_z^{(\pi)}$ by itself is not immediately useful, as the true system under consideration is really $f^{(\pi)}$; $f_z^{(\pi)}$ is an artificial construct used for computational efficiency. 
Hence, the main problem we study in this work is how to \emph{transfer} desirable verified properties from the latent system $f_z^{(\pi)}$ over to the actual system $f^{(\pi)}$, allowing us to obtain rigorous guarantees on the latter system, while paying a computational burden in the dimensionality of the former.
%while paying the computational burden of the dimensionality of the former.

\section{Measures of Approximate Conjugacy}
\label{sec:approximate_conjugacy}

A key part of our results involves quantifying discrepancies between the latent $f_z^{(\pi)}$ and
actual $f^{(\pi)}$ closed loop models.
One challenge which arises in defining a notion of error is that the outputs of both models do not have the same dimension ($n_z$ vs. $n_x$). In order to place both dynamics into the same
space, there are two possible paths:
we can compare the dynamics evolution in the latent space (\emph{forward} path), or we can compare the dynamics evolution in the original space (\emph{backward} path). 
Based on these two possibilities, we introduce
the following definitions of error.
% , both for continuous and discrete time.
We start with our continuous-time definition.
\begin{df}[Conjugate Dynamics Error -- Continuous Time]
\label{def:continuous_time_dynamics_error}
Given a set $\mathcal{D}_x\subseteq \mathbb{R}^{n_x}$, we say that the tuple $(f_z, \mathcal{D}_x)$ is \emph{$\gamma$-forward-conjugate} in continuous-time if:
\begin{equation}
    \sup_{x\in\mathcal{D}_x}\left\|\frac{\partial E}{\partial x}(x)f^{(\pi)}(x)-f_z^{(\pi)}(E(x))\right\|\leq\gamma.\label{encoder_form_error}
\end{equation}
On the other hand, we say that the tuple
$(f_z, \mathcal{D}_x)$ is \emph{$\gamma$-backward-conjugate} in continuous-time if
$\frac{\partial E}{\partial x}(x)$ has full row-rank for every $x \in \mathcal{D}_x$, and if:
\begin{equation}
    \sup_{x\in\mathcal{D}_x} \left\|f^{(\pi)}(x)-\frac{\partial E}{\partial x}(x)^{\dagger}f_z^{(\pi)}(E(x))\right\|\leq\gamma.\label{decoder_form_error}
\end{equation}
\end{df}
% \textcolor{red}{TODO: write some intuition for both losses.}
% \textcolor{red}{TODO: connect $\frac{\partial E}{\partial x}^{\dag}$ to the decoder's Jacobian.} 
Intuitively, the $\gamma$-\textit{forward conjugacy} condition ensures that the true dynamics at a point $x$, once encoded, stay close to the prediction of the latent space dynamics. Conversely, the $\gamma$-\textit{backward conjugacy} condition requires that the predicted dynamics at a point $z$ in the latent space, once decoded, faithfully approximate the true dynamics. Moreover, when we have that $E \circ D \approx \mathrm{Id}$, applying the chain rule gives
$$
    \frac{\partial(E \circ D)}{\partial z}(z)=\frac{\partial E}{\partial x}(x) \frac{\partial D}{\partial z}(z) \approx I_{n_z}\!=\!\frac{\partial E}{\partial x}(x)\left(\frac{\partial E}{\partial x}(x)\right)^{\dagger}
$$
Interpreting $x=D(z)$, this means that $\frac{\partial D}{\partial z}(z)$ acts almost as the pseudo-inverse, i.e., $\frac{\partial D}{\partial z}(z) \approx\left(\frac{\partial E}{\partial x}(x)\right)^{\dagger}$,
hence \eqref{decoder_form_error} can be interpreted as involving the Jacobian of the decoder.
%
% we can then use the Jacobian of the decoder $\partial D(z\coloneqq E(x))$ to approximate the Moore-Penrose inverse of the encoder jacobian $(\partial E(x))^\dagger$ because of the following application of chain-rule.
% \begin{equation*}
%     \frac{\partial E\circ D}{\partial z} = \frac{\partial E}{\partial x}(D({z})) \frac{\partial D}{\partial z}({z})\approx \frac{\partial E}{\partial x}(x)\frac{\partial E}{\partial x}(x)^{\dagger} \approx I_{n_z}
% \end{equation*}

We have the following similar definition for discrete-time.
\begin{df}[Conjugate Dynamics Error -- Discrete Time]
\label{def:discrete_time_dynamics_error}
Given a set $\mathcal{D}_x\subseteq \mathbb{R}^{n_x}$, we say that the tuple $(f_z, \mathcal{D}_x)$ is \emph{$\gamma$-forward-conjugate} in discrete-time if:
\begin{equation}
    \sup_{x\in\mathcal{D}_x}\left\|(E\circ f^{(\pi)})(x)-(f_z^{(\pi)}\circ E)(x)\right\|\leq\gamma.\label{encoder_form_error_discrete}
\end{equation}
On the other hand, we say that the tuple
$(f_z, \mathcal{D}_x)$ is \emph{$\gamma$-backward-conjugate} in discrete-time if $E$ has a right inverse $E^{-1}$ such that $(E\circ E^{-1})(z)=z$ for every $z \in E(\mathcal{D}_x)$, and:
\begin{equation}
    \sup_{x\in\mathcal{D}_x} \left\|f^{(\pi)}(x)-(E^{-1}\circ f_z^{(\pi)}\circ E)(x)\right\|\leq\gamma.\label{decoder_form_error_discrete}
\end{equation}
\end{df}

We briefly remark on the connections of our conjugacy definitions to 
existing literature. First, we remark that
%
%
% \textcolor{red}{TODO: connect with the SINDy work of brunton \url{https://www.pnas.org/doi/10.1073/pnas.1517384113}\url{https://arxiv.org/pdf/1904.02107}}
our continuous-time conditions (\Cref{def:continuous_time_dynamics_error}) 
reflect the two dynamics losses in SINDy~\cite{champion2019sindy}, where the authors learn an encoder $z=E(x)$ and decoder $x\approx D(z)$, together with a sparse latent dynamics model to represent the continuous-time derivative $\dot{z}$. 
Furthermore, our discrete-time conditions (\Cref{def:discrete_time_dynamics_error})
can be considered approximate versions of the invariance conditions for reduced order models
presented in \cite{szalai2023data}.
In particular, \eqref{encoder_form_error_discrete} mirrors the notion of
\emph{invariant foliations},
whereas \eqref{decoder_form_error_discrete} mirrors the notion of \emph{invariant manifolds}. Finally, we note the definition of \emph{approximately related} vector fields given in \cite{tabuada2008approximate} 
can be interpreted as a forward-conjugacy condition applied 
to trajectories of the dynamical systems in question, and
hence sits in between \eqref{encoder_form_error}
and \eqref{encoder_form_error_discrete}.

\section{Latent Lyapunov Functions}
\label{sec:lyapunov}

% Our main results are conditions on the encoding and decoding maps $E$ and $D$, and latent dynamics $f_z$, that, if satisfied, guarantee that stable control inputs for the latent system (i.e. control inputs which satisfy Lyapunov inequalities) are also stable when applied to the original system $\dot{x}=f(x,u)$ or $x_{t+1}=f(x_t,u_t)$. We assume that $f$ and $f_z$ share the same action space $\mathcal{U}\subseteq\mathbb{R}^{n_u}$. 
Our first set of results involve transfer lemmas 
for asymptotic stability induced by Lyapunov functions in the latent space.
We first introduce our continuous time results, followed by their discrete-time counterparts.

\subsection{Continuous Time Results for Lyapunov Functions}
\label{sec:lyapunov:continuous}

% Consider a continuous time system $\dot{x}=f(x,u)$ where $f:\mathbb{R}^{n_x}\times\mathbb{R}^{n_u}\rightarrow\mathbb{R}^{n_x}$, an encoding map $E:\mathbb{R}^{n_x}\rightarrow\mathbb{R}^{n_z}$, a decoding map $D:\mathbb{R}^{n_z}\rightarrow\mathbb{R}^{n_x}$, and a latent dynamics model $\dot{z}=f_z(z,u)$ where $f_z:\mathbb{R}^{n_z}\times\mathbb{R}^{n_u}\rightarrow\mathbb{R}^{n_z}$. We assume that $n_z<n_x$. For a latent feedback controller $\pi:\mathbb{R}^{n_z}\rightarrow\mathbb{R}^{n_u}$, define $f^{(\pi)}(x)=f(x,(\pi\circ E)(x))$ and $f_z^{(\pi)}(z)=f_z(z, \pi(z))$. Before introducing our results, we first define a specific class of Lyapunov function which we will analyze.

For what follows, 
we suppose that the origin is a fixed point of the zero-input continuous dynamics (i.e., that $\dot{x} = f(0,0) = 0$),
and that the encoder is continuously differentiable and origin-preserving, i.e., such that $E(0)=0$.
We fix an open set $\mathcal{D}_z \subseteq \R^{n_z}$ which contains zero, and define
$\mathcal{D}_x := E^{-1}(\mathcal{D}_z) = \{ x \in \R^{n_x} \mid E(x) \in \mathcal{D}_z \}$.
Observe that under our assumptions, the set $\mathcal{D}_x$ is open as well and contains the origin. We start our discussion by defining the notion of a (exponentially stable) Lyapunov function for the continuous time latent dynamics.
\begin{df}[Continuous Time Lyapunov Function]\label{continuous_time_lyapunov_def} 
% Let $\mathcal{D}_z \subseteq \R^{n_z}$ be an open set containing zero.
Let $\rho > 0$.
We call a function $V : \mathcal{D}_z \mapsto \R_{\geq 0}$ a $\rho$-Lyapunov function if 
(i) $V(0) = 0$ and $V(z) > 0$ for all $z \in \mathcal{D}_z \setminus \{0\}$, and
(ii) $\dot{V}(z)=\langle\nabla V(z),f_z^{(\pi)}(z)\rangle\leq-\rho V(z)$ for all $z \in \mathcal{D}_z$.
%
% Let $V:\mathcal{D}_z\subset\mathbb{R}^{n_z}\rightarrow\mathbb{R}$ be a function that satisfies $V(z)>0$ and $\dot{V}(z)=\langle\nabla V(z),f^{(\pi)}(z)\rangle\leq-\rho V(z)$, for $\rho>0$ and dynamics $\dot{z}=f^{(\pi)}(z)$, for all $z\in\mathcal{D}_z\setminus 0$, and also $\dot{V}(0)=V(0)=0$. Then we call $V(z)$ a Lyapunov function on domain $\mathcal{D}_z$. If the domain of $V$ is in the output space of the encoder $E$, then we call $V$ a continuous time latent Lyapunov function. 
\end{df}

We note that a natural way to design a latent space controller $\pi(z)$ is by using the framework of control Lyapunov functions. Here, $\pi(z)$ can be obtained analytically 
(e.g., Sontag's universal formula)
or via the optimization problem $\pi(z) =\argmin_{u \in \mathcal{U}} \langle \nabla V(z), f_z(z, u) \rangle$. We remark that our results are agnostic to the actual control design procedure.

Next, we describe a key lemma that characterizes the relationship between upper bounds on the conjugate dynamics error (cf.~\Cref{def:continuous_time_dynamics_error}) and its 
implication on the time derivative of the 
%\textcolor{red}{EDIT HERE} \emph{almost} Lyapunov 
function
%~\cite{liu2020almost} 
$\overline{V} = V \circ E$.
%that guarantees practical stability with respect to the set $E^{-1}(0)$. 
%Further discussion follows after the lemma.

\begin{lmm}[Lyapunov Condition (Continuous Time)]\label{continuous_lyapunov_lemma}
% Let $\mathcal{D}_z \subseteq \R^{n_z}$ be an open set containing zero, and 
Let $\dot{x}=f^{(\pi)}(x)$ be continuous on $\mathcal{D}_x$ and
$V : \mathcal{D}_z \mapsto \R_{\geq 0}$ be a $\rho$-Lyapunov function (cf. \Cref{continuous_time_lyapunov_def}).
% Set $\mathcal{D}_x := E^{-1}(\mathcal{D}_z) = \{ x \in \R^{n_x} \mid \exists z \in \mathcal{D}_z \text{ s.t. } E(x) = z \}$.
Suppose either that 
(i) $V$ is $L$-Lipschitz on $\mathcal{D}_z$ and $(f_z, \mathcal{D}_x)$ is $\gamma$-forward-conjugate (cf.~\Cref{encoder_form_error}),
or (ii) $V \circ E$ is $L$-Lipschitz on $\mathcal{D}_x$ and 
$(f_z, \mathcal{D}_x)$ is $\gamma$-backward-conjugate (cf.~\Cref{decoder_form_error}).
Then, for all $x \in \mathcal{D}_x$, it holds that
\begin{equation}\label{continuous_time_lyapunov}
    \dot{\overline{V}}(x)+\rho \overline{V}(x)\leq L\gamma,
\end{equation}
where the function $\overline{V} := V \circ E$ is defined over the domain $\mathcal{D}_x$, i.e., $\overline{V}:\mathcal{D}_x\mapsto \R_{\geq 0}$. 
\end{lmm}
%
% \begin{lmm}[Lyapunov Inequality for Approximately-Conjugate Dynamics -- Continuous Time]\label{continuous_lyapunov_lemma}
% Consider encoding map $E:\mathbb{R}^{n_x}\rightarrow\mathbb{R}^{n_z}$, original dynamics $f:\mathbb{R}^{n_x}\times\mathbb{R}^{n_u}\rightarrow\mathbb{R}^{n_x}$, latent dynamics $f_z:\mathbb{R}^{n_z}\times\mathbb{R}^{n_u}\rightarrow\mathbb{R}^{n_z}$, and latent feedback controller $\pi:\mathbb{R}^{n_z}\rightarrow\mathbb{R}^{n_u}$. Let $V:\mathcal{D}_z\subset\mathbb{R}^{n_z}\rightarrow\mathbb{R}$ be an $L$-Lipschitz function that satisfies $\langle \nabla V(z),f^{(\pi)}_z(z)\rangle+\rho V(z)\leq0$ where $\rho>0$, for all $z\in\mathcal{D}_z$. With $\mathcal{D}_x=E^{-1}(\mathcal{D}_z)$, if either the conjugate dynamics error of the encoding or inverse encoding, \cref{encoder_form_error} or \cref{decoder_form_error}, are upper-bounded by $\gamma$, then
% \begin{equation}\label{continuous_time_lyapunov}
%     \dot{\overline{V}}(x)+\rho \overline{V}(x)\leq L\gamma
% \end{equation}
% for $\overline{V}=V\circ E$. If \cref{decoder_form_error} is the tighter bound, then assume $V\circ E$ is $L$-Lipschitz.
% \end{lmm}
%
\toggle{
\begin{proof} 
We first show that a $\rho$-Lyapunov function $V$ under condition (i) implies that \Cref{continuous_time_lyapunov} holds for all $x \in \mathcal{D}_x$. For this, we start by computing $\dot{\overline{V}}(x)$ as
\begin{align}
    \dot{\overline{V}}(x)&\overset{(a)}{=}\left\langle\frac{\partial E}{\partial x}(x)^T\nabla V(E(x)),f(x,(\pi\circ E)(x))\right\rangle \nonumber\\
    &\overset{(b)}{=}\bigg\langle\nabla V(E(x)),\frac{\partial E}{\partial x}(x)f(x,\pi (E(x)))\nonumber\\
    &-f_z(E(x),\pi( E(x)))\bigg\rangle\\
    &+\bigg\langle\nabla V(E(x)),f_z(E(x),\pi( E(x)))\bigg\rangle \label{eq:V_dot_bar}
\end{align}
where $(a)$ follows from the chain rule and $(b)$ is simple algebraic manipulation. From \Cref{eq:V_dot_bar}, we obtain the following sequence of inequalities for $\dot{\overline{V}}(x)$:
\begin{align}
    \dot{\overline{V}}(x) &\overset{(c)}{\leq}\bigg\|\nabla V(E(x))\bigg\|\cdot\bigg\|\frac{\partial E}{\partial x}(x)f(x,(\pi\circ E)(x)) \nonumber\\
    &-f_z(E(x),(\pi\circ E)(x))\bigg\|\nonumber\\
    &+\bigg\langle\nabla V(E(x)),f_z(E(x),\pi(E(x)))\bigg\rangle \label{eq:11}\\
    &\overset{(d)}{\leq} L \gamma +\bigg\langle\nabla V(E(x)),f_z(E(x),\pi(E(x)))\bigg\rangle \label{eq:V_dot_bar_}
\end{align}
where $(c)$ follows by the Cauchy-Schwarz inequality and $(d)$ follows by $L$-Lipschitz continuity of $V$ and since $(f_z, \mathcal{D}_x)$ is $\gamma$-forward-conjugate. Because \Cref{eq:V_dot_bar_} holds
for all $x \in \mathcal{D}_x$, $V$ is a $\rho$-Lyapunov function, and $z=E(x)$, we  conclude  $\dot{\overline{V}}(x)\leq L\gamma-\rho\overline{V}(x)$ for all $x \in \mathcal{D}_x$.

Similarly, we next show that a $\rho$-Lyapunov function $V$ together with condition (ii) implies that \Cref{continuous_time_lyapunov} holds for all $x \in \mathcal{D}_x$. For brevity, we now use $\partial E(x):=\frac{\partial E}{\partial x}(x)$. Our starting point is \Cref{eq:V_dot_bar} from which we get
\begin{align*}
\dot{\overline{V}}(x)&=\bigg\langle\nabla V(E(x)),\partial{E}(x)f^{(\pi)}(x)\\
&\quad-\partial E(x)\partial E(x)^T(\partial E(x)\partial E(x)^T)^{-1}f_z^{(\pi)}(E(x))\bigg\rangle\\
&\quad+\bigg\langle\nabla V(E(x)),f_z^{(\pi)}(E(x))\bigg\rangle
\end{align*}
by inserting an identity matrix. From here, using simple manipulations, we can further derive 
\begin{align*}
\dot{\overline{V}}(x)&=\bigg\langle\nabla V(E(x)),\partial E(x)\bigg(f^{(\pi)}(x)\\
&\quad -\partial E(x)^T(\partial E(x)\partial E(x)^T)^{-1} f_z^{(\pi)}(E(x))\bigg)\bigg\rangle\\
&\quad+\bigg\langle\nabla V(E(x)),f_z^{(\pi)}(E(x))\bigg\rangle\\
&=\bigg\langle\partial E(x)^T\nabla V(E(x)), f^{(\pi)}(x)-\partial E(x)^\dagger f_z^{(\pi)}(E(x))\bigg\rangle\\
&\quad+\bigg\langle\nabla V(E(x)),f_z^{(\pi)}(E(x))\bigg\rangle\\
&=\bigg\langle\nabla\overline{V}(x), f^{(\pi)}(x)-\partial E(x)^\dagger f_z^{(\pi)}(E(x))\bigg\rangle\\
&\quad+\bigg\langle\nabla V(E(x)),f_z^{(\pi)}(E(x))\bigg\rangle.
\end{align*}
We can now expand this expression using the Cauchy Schwarz inequality as
\begin{align}
\dot{\overline{V}}(x)&\leq\big\|\nabla\overline{V}(x)\big\|\cdot \big\|f^{(\pi)}(x)-\partial E(x)^\dagger f_z^{(\pi)}(E(x))\big\|\nonumber\\
&\quad+\bigg\langle\nabla V(E(x)),f_z^{(\pi)}(E(x))\bigg\rangle\nonumber\\
&\overset{(e)}{\leq} L\gamma+\bigg\langle\nabla V(E(x)),f_z^{(\pi)}(E(x))\bigg\rangle \label{eq:llll} 
\end{align}
where $(e)$ follows by $L$-Lipschitz continuity of $V\circ E$ and since $(f_z, \mathcal{D}_x)$ is $\gamma$-backward-conjugate. Since \Cref{eq:llll} holds for all $x\in\mathcal{D}_x$, $V$ is a $\rho$-Lyapunov function, and $z=E(x)$, we conclude  $\dot{\overline{V}}(x)\leq L\gamma-\rho\overline{V}(x)$ for all $x \in \mathcal{D}_x$. 
\end{proof}
}

\toggle{
\begin{rem}[Significance of $\mathcal{D}_x$]
    The bounds obtained in \Cref{continuous_lyapunov_lemma} necessitate that the supremum in the definitions of the forward and backward conjugacy in \cref{encoder_form_error,decoder_form_error,encoder_form_error_discrete,decoder_form_error_discrete} be computed over $\mathcal{D}_x=E^{-1}(\mathcal{D}_z)$ to account for all the states the system may encounter when operating in $\mathcal{D}_z$, yielding the upper bounds in \cref{eq:V_dot_bar_} and \cref{eq:llll}. This is done instead of computing $\gamma$
    over the simpler set given by the image of $\mathcal{D}_z$ under the decoder, 
    i.e., $D(\mathcal{D}_z)$.
    % $D(\mathcal{D}_z)$ using the decoder. 
    The reason for this is that $D(E(x))$ may not be equal to $x$. This can happen for two reasons: (1) Due to approximation errors (e.g., introduced by learning), $D(E(x))$ may not be contained in the inverse image of $E(x)$, i.e. $E(D(E(x)))\neq E(x)$. (2) The decoder $D$ is not a set-valued function, whereas there may be multiple states $x$ that map to $z=E(x)$ through $E$.
    %To see this, observe that the inequalities obtained in \cref{eq:V_dot_bar_} and \cref{eq:llll} must hold over all $x\in\mathcal{D}_x$. 
    Consider two points $x, x'\in \mathcal{D}_x$ such that $x\neq x'$ and $E(x)=E(x')$. The decoder $D$ will
    %, at best, 
    only be able to recover either $x$ or $x'$, i.e. $D(z)=x$ or $D(z)=x'$, but not both.
    %(in fact, for learned or approximate $D$, $D(z)$ may not even be an element of $E^{-1}(z)$). 
    In general, $f(x,u) \neq f(x',u)$ and thus e.g., for 
    the forward conjugacy error $\gamma(x):=\|\frac{\partial E}{\partial x}(x)f^{(\pi)}(x)-f_z^{(\pi)}(E(x))\|$, 
    we generally have $\gamma(x)\neq\gamma(x')$. As a result, we must use the bound $\sup_{x\in\mathcal{D}_x}\gamma(x)$, rather than the simpler $\sup_{z\in\mathcal{D}_z}\gamma(D(z))$.
    A similar line of reasoning also holds for the 
    backward conjugacy error.
\end{rem}}

\color{black}

We next show that, if the hypotheses in the preceding lemma hold, we can guarantee the existence of a set $\mathcal{V}_\alpha\subseteq \mathcal{D}_x$ defined via the level-sets of $\overline{V}$ that is forward invariant and attractive for the original dynamics $\dot{x}=f^{(\pi)}(x)$.

%\begin{as}[Positive Definite Encoder]\label{positive_definite_encoder}
%Assume that $E:\mathbb{R}^{n_x}\rightarrow\mathbb{R}^{n_z}$ is continuous and positive definite on the domain $\mathcal{D}_x$, i.e., that $E(0)=0$ and  $E(x)\neq0$ for $x\in\mathcal{D}_x$.
%\end{as}

% \textcolor{red}{TODO: I think the proposition below needs a bit of work.
% I think we want to show two things:
% \begin{enumerate}
%     \item Suppose $x(0) \in \mathcal{D}_x$ starts such that $x(0) \not\in \mathcal{V}_{L\gamma/\rho}$.
%     Then, want to show something like $\bar{V}(x(t)) - L\gamma/\rho \leq \rho^{t} \bar{V}(x(0))$.
%     \item Now suppose that $x(t)$ enters
%     $\mathcal{V}_{L\gamma/\rho}$ at some $t$, then
%     $x(t') \in \mathcal{V}_{L\gamma/\rho}$ for all $t' \geq t$.
% \end{enumerate}
% I believe both could probably be shown with a barrier function type argument. What the current proof below establishes is that for $V' = \bar{V} - L\gamma/\rho$ we have the differential inequality $\dot{V}' \leq - \rho V$. I think one of those comparison lemma type arguments done in CBF would then conclude both points; however what is currently written is not sufficient.
% }

\begin{prp}[Stability Guarantee (Continuous Time)]\label{continuous_lyapunov_proposition} Let $\dot{x}=f^{(\pi)}(x)$  and $V:\mathcal{D}_z\mapsto \mathbb{R}_{\ge 0}$ satisfy the hypothesis of Lemma \ref{continuous_lyapunov_lemma},
and suppose that $\dot{x} = f^{(\pi)}(x)$ is forward complete, i.e., solutions $x:\mathcal{I}\mapsto \mathcal{D}_x$ to $\dot{x} = f^{(\pi)}(x)$ are such that $\mathcal{I}= \mathbb{R}_{\ge 0}$.
%Furthermore, assume that $\mathcal{D}_z$ is a compact set. 
Define the sub-level sets $\mathcal{V}_\alpha=\{x\in\mathbb{R}^{n_x} \mid \overline{V}(x)\leq\alpha\}$, and critical radius $\alpha_0=\sup\{\alpha> 0 \mid \mathcal{V}_\alpha \subseteq \mathcal{D}_x\}$. %Assume that the sub-level set $\mathcal{V}_{\alpha_0}$ is compact. 
% \textcolor{blue}{Assume that the system $f^{(\pi)}(x)$ in \cref{eq:system} is forward complete.} 
%under bounded inputs $\sup_{t\geq0}\|u_t\|<\infty$, and that the controller $\pi(z)$ is continuous over $\mathcal{D}_z$. 
For an initial condition $x_0\in\mathcal{V}_{\alpha_0}$, we have that for all $t\in\mathbb{R}_{\ge 0}$,
\begin{align}
\overline{V}(x_t)\le (\overline{V}(x_0)-L\gamma/\rho)\exp(-\rho t)+L\gamma/\rho.
\end{align}
Thus, the set  $\mathcal{V}_\alpha$ is forward invariant for any $\alpha\in [L\gamma/\rho,\alpha_0]$ and the set $\mathcal{V}_{L\gamma/\rho}$ is attractive from $\mathcal{V}_{\alpha_0}\setminus \mathcal{V}_{L\gamma/\rho}$.
\end{prp}
\toggle{
\begin{proof}
Recall from \cref{continuous_lyapunov_lemma} that 
 $\dot{\overline{V}}(x)\leq L\gamma-\rho \overline{V}(x)$ holds for all $x \in \mathcal{D}_x$. Let now $x:\mathcal{I}\mapsto \mathcal{D}_x$ be a solution to the initial value problem of \eqref{eq:system} with initial condition $x_0\in \mathcal{V}_{\alpha_0}$ and where $\mathcal{I}\subseteq \mathbb{R}_{\ge 0}$ is the maximum domain of  $x$. Consequently, we know that $\dot{\overline{V}}(x_t)\leq L\gamma-\rho \overline{V}(x_t)$ for all $t\in\mathcal{I}$. We next apply the comparison lemma, see \cite[Theorem 3.3]{khalil2002nonlinear}. Therefore, consider the initial value problem 
 \begin{align*}
     \dot{w}_t = L\gamma-\rho w_t \text{ with }w_0:=\bar{V}(x_0).
 \end{align*}
 One can verify that the solution $w:\mathbb{R}_{\ge 0} \mapsto \mathbb{R}$ to this initial value problem is $w_t:=(w_0-L\gamma/\rho)\exp(-\rho t)+L\gamma/\rho$. Using the comparison lemma, we  get that  $\overline{V}(x_t)\le (\bar{V}(x_0)-L\gamma/\rho)\exp(-\rho t)+L\gamma/\rho$ for all $t\in\mathcal{I}$.
 %What remains to be shown is that the solution $x$ is complete, i.e., that $\mathcal{I}=\mathbb{R}_{\ge 0}$. %In that regard, recall that $\mathcal{V}_{\alpha_0}$ is assumed to be a compact set. By  \cite[Theorem 3.3]{khalil2002nonlinear}, 
Recall our assumption that $\dot x=f^{(\pi)}(x)$ is forward complete. Thus, $\mathcal{I}=\mathbb{R}_{\ge 0}$ and the proof is complete.
%Recall that $\pi(z)$ is continuous for all $z\in\mathcal{D}_z$ where $\mathcal{D}_z$ is a compact set. By the extreme value theorem, it follows that $\pi(z)$ is bounded for all $z\in\mathcal{D}_z$. By definition of $\mathcal{D}_x$, it then follows that $(\pi\circ  E)(x)$ is bounded for all $x\in\mathcal{D}_x$. Furthermore, we have that the system $f(x,u)$ is forward complete for bounded control inputs $\sup_{t\geq0}\|u_t\|<\infty$. As the control input is $u_t=(\pi\circ E)(x_t)$, it follows that $\sup_{t\geq0}\|u_t\|<\infty$ so that we  conclude  $\mathcal{I}=\mathbb{R}_{\ge 0}$.}
\end{proof}
}

\toggle{
\begin{rem}[Forward Completeness]
    In \Cref{continuous_lyapunov_proposition},
    we make the assumption that the solution to $\dot{x} = f^{(\pi)}(x)$ is forward complete.
    One way to satisfy such an assumption is to assume that the set $\mathcal{D}_x$ is compact; this would however greatly restrict the form of $(E, D)$ as e.g., $\mathcal{D}_x$ is unbounded for linear encoders.
    To avoid imposing such a restrictive
    compactness assumption, in Appendix~\ref{app:cartpole_forward_completeness}
    we give a self-contained
    proof of a general result (\Cref{prp:forward_completeness}) that shows
    for systems where the Hamiltonian equals the total energy (e.g., a broad class of mechanical systems), bounded control inputs imply forward completeness.
    % Rather than assuming compactness of $\mathcal{D}_x$ in order to obtain forward completeness in \Cref{continuous_lyapunov_proposition}, which would greatly restrict the form of $E, D$ (e.g. $E^{-1}(0)$ is unbounded for linear encoders), we instead directly assume that the system $f^{(\pi)}(x)$ is forward complete, i.e. solutions exist for all time. In \Cref{prp:forward_completeness} (cf. \cref{app:cartpole_forward_completeness}), we show that, for systems where the Hamiltonian equals the total energy (e.g. mechanical systems), bounded control inputs implies forward completeness. We apply this result in \Cref{app:cartpole_forward_completeness} to show that the cartpole system from \Cref{sec:experiments:cartpole} is forward complete under bounded control inputs.\\ 
    % \textcolor{red}{cbf survey papers? iss stability paper?} 
\end{rem}}
%\color{black}
\begin{rem}[Relation to Almost Lyapunov Functions]
We highlight the relation between our use of \cref{continuous_lyapunov_lemma}'s uniform bound on the time derivative of $\overline{V}=V\circ E$ in \Cref{continuous_lyapunov_proposition}, and the concept of ``almost Lyapunov functions" in~\cite{liu2020almost}. In \cref{continuous_lyapunov_proposition}, we establish an implication between uniform dynamics error and the forward-invariant and attractive sets of $\overline{V}$. In a similar vein, almost Lyapunov functions are allowed to violate the decrease condition over a subset of their domain, yielding convergence to a neighborhood around the equilibrium.
\end{rem}

  We illustrate the sets from \Cref{continuous_lyapunov_proposition} in \Cref{fig:latent_safe_set}. However, recall that our initial goal was to show that the origin of $\dot{x} = f^{(\pi)}(x)$ is stable. In contrast, \Cref{continuous_lyapunov_proposition}  shows that $\mathcal{V}_{L\gamma/\rho}$ is attractive from $\mathcal{V}_{\alpha_0}\setminus \mathcal{V}_{L\gamma/\rho}$. 
   A natural question is what this set attraction implies about
   the behavior of $x_t$.
   To further analyze this behavior, let us first suppose there
   exists a class $\mathcal{K}$-function $\kappa : \R_{\geq 0} \mapsto \R_{\geq 0}$ such that $\kappa(\|z\|) \leq V(z)$ for all $z \in \R^{n_z}$, from which by \Cref{continuous_lyapunov_proposition} it follows that
    %   
   % If we also impose that $E:\mathbb{R}^{n_x} \mapsto \mathbb{R}^{n_z}$ is such that there exists a class $\mathcal{K}$-function $\kappa:\mathbb{R}_{\ge 0} \mapsto \mathbb{R}_{\ge 0}$ such that $\kappa(\|x\|)\le \bar{V}(x)$, then it holds that
   \begin{align*}
       \|E(x_t)\|\le \kappa^{-1}\big((\overline{V}(x_0)-L\gamma/\rho)\exp(-\rho t)+L\gamma/\rho\big)
   \end{align*}
   for all $t \geq 0$. This then implies that 
\begin{align}
    \limsup_{t\to \infty}\|E(x_t)\|\le \kappa^{-1}(L\gamma/\rho), \label{eq:encoder_convergence}
\end{align}
i.e., that $\|E(x_t)\|$ converges to a region of size $\kappa^{-1}(L\gamma/\rho)$ around the origin.
   %
   % so that  we achieve practical stability. Note that we can achieve $\kappa(\|x\|)\le \bar{V}(x)$ if we impose that (i) $E(0)=0$, and (ii) $E$ is monotonically increasing with $\|x\|$.
   % \textcolor{red}{(a) we already impose $E(0)=0$ earlier in the text, and (b) maybe say radially unbounded or just be explicitly $E(x) \to \infty$ as $\|x\| \to\infty$? }
%\end{rem}
%
We now turn to understanding the behavior of $x_t$ from $E(x_t)$. 
Ideally, one could establish some condition such that $E(x_t) \to 0$ implies $x_t \to 0$. 
Unfortunately, this is generally not true when $n_z < n_x$. To see this, a first-order Taylor expansion of the encoder $E(x)$ yields 
$E(x_t) \approx E(0) + \partial E(x_t) x_t = \partial E(x_t) x_t$,
since $E(0) = 0$. However, the Jacobian matrix $\partial E(x) \in \R^{n_z \times n_x}$ necessarily has a non-trivial nullspace whenever $n_z < n_x$, which means it is possible to choose a sequence of 
vectors $\{ v_k \}_{k \geq 0}$ such that $E(v_k) \to 0$ but $\| v_k \| = 1$ for all $k$. Hence, the statement \eqref{eq:encoder_convergence} cannot be refined in general.
This argument illustrates the necessity of carefully designing 
the encoder $E(x)$ so that small values of $\| E(x) \|$ are meaningful for the stability of the 
control task, when $x$ is restricted to a relevant sub-manifold.
% In \Cref{sec:learning}, we explore the various design decision involved in designing learning algorithms for $E(x)$.
\begin{rem}[Practical Stability]
    To summarize the previous discussion, we reiterate that even in the case of zero conjugacy error, the stability guarantee obtained is practical stability \cite{moreau2002practical}, i.e., stability to the set $E^{-1}(0)$. In general, because $E$ will most likely have a nontrivial kernel, the set $E^{-1}(0)$ is not a singleton set.
    %    
    % rather than a point. Because $E$ will most likely have a nontrivial kernel, we obtain stability to the set $E^{-1}(0)$. 
    Thus, the conjugacy error will induce forward-invariant and attractive level sets of $\overline{V}$ around the zero level-set given by $E^{-1}(0)$.
\end{rem}

We observe that both forward and backwards
conjugate losses (cf.~\Cref{def:continuous_time_dynamics_error})
enable stability guarantees to transfer over from the latent to the original system. 
In practice, this provides additional flexibility when learning latent dynamics, as there are two valid reconstruction losses which can be used during training. We leave a more in-depth investigation into the implications of this finding 
as future work.
We also remark that the decoder map $D$ plays no direct role in \Cref{continuous_lyapunov_lemma} and \Cref{continuous_lyapunov_proposition}. Instead, it plays the indirect role of regularizing $E$ during training, e.g., preventing the encoder from learning trivial latent representations such as mapping all vectors to the origin of the latent space.

% \textcolor{red}{TODO: is the part about our experiments showing improved performance using these losses still true?}
% Further observe that \Cref{continuous_lyapunov_lemma} introduced \Cref{encoder_form_error} and \Cref{decoder_form_error}, a pair of conjugate dynamics errors that, if sufficiently small, allow systems to inherit forward invariance and attractivity properties as shown in \Cref{continuous_lyapunov_proposition}. In our experiments, we show that including these errors as loss functions during training improves the performance of our learned latent dynamics, and encoder-decoder pair, when $E$, $D$, and $f_z$ are parameterized as neural networks. Also note that the decoding map $D$ plays no role in \Cref{continuous_lyapunov_lemma} and \Cref{continuous_lyapunov_proposition}. It serves the purpose of regularizing $E$ during training, preventing the encoder from learning trivial latent representations such as mapping all vectors to the origin of the latent space. 

\subsection{Discrete Time Results for Lyapunov Functions}
\label{sec:lyapunov:discrete}

% Consider a discrete time system $x_{t+1}=f(x_t,u_t)$ where $f:\mathbb{R}^{n_x}\times\mathbb{R}^{n_u}\rightarrow\mathbb{R}^{n_x}$. Like in the continuous time case, we consider an encoding map $E:\mathbb{R}^{n_x}\rightarrow\mathbb{R}^{n_z}$, a decoding map $D:\mathbb{R}^{n_z}\rightarrow\mathbb{R}^{n_x}$, and latent dynamics $\dot{z}=f_z(z,u)$, where $f_z:\mathbb{R}^{n_z}\times\mathbb{R}^{n_u}\rightarrow\mathbb{R}^{n_z}$. Consider a latent feedback controller $\pi:\mathbb{R}^{n_z}\rightarrow\mathbb{R}^{n_u}$ and define $f^{(\pi)}(x)=f(x,(E\circ\pi)(x))$ and $f_z^{(\pi)}(z)=f_z(z,\pi(z))$. We first define the class of Lyapunov function we are interested in.

We now turn to discrete time systems. Here, we again suppose that the origin is a fixed
point of the zero-input discrete dynamics (i.e., that $x_{t+1} = f(0, 0)=0$).  In this section, we carry over the definition of the sets $\mathcal{D}_z$ and $\mathcal{D}_x$ from \Cref{sec:lyapunov:continuous}.
We start with the definition of a discrete-time (exponentially stable) Lyapunov function.

\begin{df}[Discrete Time Lyapunov Function]\label{discrete_time_lyapunov_def}
Let $\rho \in [0, 1)$.
We call a function $V : \mathcal{D}_z \mapsto \R_{\geq 0}$ a $\rho$-discrete-Lyapunov function if 
(i) $V(0) = 0$ and $V(z) > 0$ for all $z \in \mathcal{D}_z \setminus \{0\}$, and
(ii) $\Delta V(z)=V(f_z^{(\pi)}(z))-\rho V(z)\leq0$ for all $z \in \mathcal{D}_z$.

% Let $V:\mathcal{D}_z\subset\mathbb{R}^{n_z}\rightarrow\mathbb{R}$ be a function that satisfies $V(z)>0$ and $\Delta V(z)=V(f^{(\pi)}(z))-\rho V(z)\leq0$, for $\rho\in(0,1]$ and dynamics $\dot{z}=f^{(\pi)}(z)$, for all $z\in\mathcal{D}_z\setminus 0$, and also $\Delta{V}(0)=V(0)=0$. Then we call $V(z)$ a Lyapunov function on domain $\mathcal{D}_z$. If the domain of $V$ is in the output space of the encoder $E$, then we call $V$ a discrete time latent Lyapunov function. 
\end{df}

As in the continuous time case, the next lemma characterizes the relationship between upper bounds on the conjugate dynamics error (cf.~\Cref{def:discrete_time_dynamics_error}) and its 
implication on the time derivative of the %\emph{almost} Lyapunov 
function $\overline{V} = V \circ E$.

\begin{lmm}[Lyapunov Condition (Discrete Time)]\label{discrete_lyapunov_lemma}
Let $x_{t+1} = f^{(\pi)}(x_t)$ and $V : \mathcal{D}_z \mapsto \R_{\geq 0}$ be a $\rho$-discrete-Lyapunov function (cf. \Cref{discrete_time_lyapunov_def}).
% Set $\mathcal{D}_x := E^{-1}(\mathcal{D}_z) = \{ x \in \R^{n_x} \mid \exists z \in \mathcal{D}_z \text{ s.t. } E(x) = z \}$.
Suppose either that 
(i) $V$ is $L$-Lipschitz on $\mathcal{D}_z$ and $(f_z, \mathcal{D}_x)$ is $\gamma$-forward-conjugate in discrete-time (cf.~\Cref{encoder_form_error_discrete}),
or (ii) $V \circ E$ is $L$-Lipschitz on $\mathcal{D}_x$ and 
$(f_z, \mathcal{D}_x)$ is $\gamma$-backward-conjugate in discrete-time (cf.~\Cref{decoder_form_error_discrete}).
Then, for all $x \in \mathcal{D}_x$, it holds that
\begin{equation}\label{discrete_time_lyapunov}
    \overline{V}(f^{(\pi)}(x))-\rho\overline{V}(x)\leq L\gamma,
\end{equation}
where the function $\overline{V} := V \circ E$ is defined over the domain $\mathcal{D}_x$, i.e., $\overline{V}:\mathcal{D}_x\mapsto \R_{\geq 0}$. 
%
% Consider encoding map $E:\mathbb{R}^{n_x}\rightarrow\mathbb{R}^{n_z}$, original dynamics $f:\mathbb{R}^{n_x}\times\mathbb{R}^{n_u}\rightarrow\mathbb{R}^{n_x}$, latent dynamics $f_z:\mathbb{R}^{n_z}\times\mathbb{R}^{n_u}\rightarrow\mathbb{R}^{n_z}$, and latent feedback controller $\pi:\mathbb{R}^{n_z}\rightarrow\mathbb{R}^{n_u}$. Let $V:\mathcal{D}_z\subset\mathbb{R}^{n_z}\rightarrow\mathbb{R}$ be an $L$-Lipschitz function that satisfies $V(f_z^{(\pi)}(z))-\rho V(z)\leq0$ where $\rho\in(0,1]$, for all $z\in\mathcal{D}_z$. With $\mathcal{D}_x=E^{-1}(\mathcal{D}_z)$, if either the conjugate dynamics error of the encoding or inverse encoding, \cref{encoder_form_error_discrete} or \cref{decoder_form_error_discrete}, are upper-bounded by $\gamma$, then
% \begin{equation}\label{discrete_time_lyapunov}
%     \overline{V}(f^{(\pi)}(x))-\rho\overline{V}(x)\leq L\gamma
% \end{equation}
% for $\overline{V}=V\circ E$. If \cref{decoder_form_error_discrete} is the tighter bound, then assume that $V\circ E$ is L-Lipschitz.
\end{lmm}
\toggle{
\begin{proof}
We first show that a $\rho$-discrete-Lyapunov function $V$ under condition (i) implies that \Cref{discrete_time_lyapunov} holds for all $x \in \mathcal{D}_x$. We start by upper bounding $\overline{V}(f^{(\pi)}(x))-\rho\overline{V}(x)$:
\begin{align*}
\overline{V}(f^{(\pi)}(x))-\rho\overline{V}(x)&=V(E( f^{(\pi)}(x)))-V(f_z^{(\pi)}( E(x)))\\ & + V(f_z^{(\pi)}(E(x)))-\rho V(E(x))\\
\leq L\gamma&+V(f_z^{(\pi)}(E(x)))-\rho V(E(x)) 
\end{align*}
where the inequality follows by $L$-Lipschitz continuity of $V$ and since $(f_z, \mathcal{D}_x)$ is $\gamma$-forward-conjugate. Because $V$ is a $\rho$-discrete-Lyapunov function and $z=E(x)$, we  conclude  $\overline{V}(f^{(\pi)}(x))-\rho\overline{V}(x)\leq L\gamma$ for all $x \in \mathcal{D}_x$. 

Similarly, we next show that a $\rho$-discrete-Lyapunov function $V$ under condition (ii) implies that \Cref{discrete_time_lyapunov} holds for all $x \in \mathcal{D}_x$. We again upper bound $\overline{V}(f^{(\pi)}(x))-\rho\overline{V}(x)$:
\begin{align*}
\overline{V}(f^{(\pi)}(x))\!-\!\rho\overline{V}(x)&=\overline{V}(f^{(\pi)}(x))-(\overline{V}\!\circ \!E^{-1}\!\circ\!f_z^{(\pi)}\!\circ\!E)(x)\\&+(\overline{V}\circ E^{-1}\circ f_z^{(\pi)} \circ E)(x)-\rho\overline{V}(x)\\
&\leq L\gamma+V(f_z^{(\pi)}(E(x)))-\rho V(E(x))
\end{align*}
where the inequality follows by $L$-Lipschitz continuity of $V\circ E$ and since $(f_z, \mathcal{D}_x)$ is $\gamma$-backward-conjugate. Because $V$ is a $\rho$-discrete-Lyapunov function and $z=E(x)$, we  conclude  $\overline{V}(f^{(\pi)}(x))-\rho\overline{V}(x)\leq L\gamma$ for all $x \in \mathcal{D}_x$. 
\end{proof}
}

As in the continuous time case, if the hypotheses in the preceding lemma hold, we show that we can guarantee the existence of a set $\mathcal{V}_\alpha\subseteq \mathcal{D}_x$ defined via the level-sets of $\overline{V}$ that is forward invariant and attractive for the original dynamics $x_{t+1}=f^{(\pi)}(x_t)$.

% \textcolor{red}{TODO: this result below also has similar issues as the cts time argument.
% I think the fix in discrete time would look like follows. As before we define
% \begin{enumerate}
%     \item First, we set
%     $$
%         \bar{V}_0 := \sup\{ v \mid \{ x \mid \bar{V}(x) \leq v + \frac{L\gamma}{1-\rho} \} \subseteq \mathcal{D}_x \}.
%     $$
%     \item Now, we use induction to prove that if $x_0 \in \{ x \mid \bar{V}(x) \leq \bar{V}_0 + \frac{L\gamma}{1-\rho} \}$, then 
%     for all $t$, we have $x_t \in \mathcal{D}_x$.
%     \item Once we have this inclusion, we can then conclude that for all $t$,
%     $$
%         \bar{V}(x_t) \leq \rho^t \bar{V}(x_0) + \frac{L\gamma}{1-\rho}.
%     $$
% \end{enumerate}
%}

\begin{prp}[Stability Guarantee (Discrete Time)]
\label{discrete_lyapunov_proposition} 
Let $x_{t+1} = f^{(\pi)}(x_t)$ and $V : \mathcal{D}_z \mapsto \R_{\geq 0}$ satisfy the hypothesis of \Cref{discrete_lyapunov_lemma}.
Define the sub-level sets $\mathcal{V}_\alpha := \{ x \mid \overline{V}(x) \leq \alpha \}$, and critical radius $\alpha_0 := \sup\{ \alpha > 0 \mid \mathcal{V}_\alpha \subseteq \mathcal{D}_x \}$. Then, the set  $\mathcal{V}_\alpha$ is forward invariant for any $\alpha\in [L\gamma/(1-\rho),\alpha_0]$. For an initial condition $x_0 \in \mathcal{V}_{\alpha_0}$,  we  have that for all $t \ge 0$,
\begin{align}
    \overline{V}(x_t) \leq \rho^t \overline{V}(x_0) + \frac{L\gamma}{1-\rho}. \label{eq:discrete_time_lyapunov_decrease}
\end{align}
% where $\bar{V} := V \circ E$.
% Furthermore, we have that the set $\mathcal{V}_0$ is forward invariant under the discrete-time dynamics $f^{(\pi)}$.
Thus, the set $\mathcal{V}_{L\gamma/(1-\rho)}$ is attractive from $\mathcal{V}_{\alpha_0}\setminus \mathcal{V}_{L\gamma/(1-\rho)}$.
\end{prp}
\toggle{
\begin{proof}
Fix $\alpha \in [L\gamma/(1-\rho), \alpha_0]$.
We first show by induction that $x_0  \in \mathcal{V}_{\alpha}$ implies $x_t \in \mathcal{V}_{\alpha}$ for all $t \geq 0$. If $x_t \in \mathcal{V}_{\alpha}$, we can upper bound $\overline{V}(x_{t+1})$ as
\begin{align*}
    \overline{V}(x_{t+1}) \overset{(a)}{\leq} \rho \overline{V}(x_t) + L \gamma \overset{(b)}{\leq} \rho \alpha + L\gamma \overset{(c)}{\leq} \alpha
\end{align*}
where $(a)$ follows from \Cref{discrete_lyapunov_lemma} and $\mathcal{V}_{\alpha} \subseteq \mathcal{D}_x$, $(b)$ holds since $x_t \in \mathcal{V}_{\alpha}$, and $(c)$ follows since $\alpha \geq L\gamma/(1-\rho)$. Consequently, $x_{t+1} \in \mathcal{V}_{\alpha}$ and $\mathcal{V}_\alpha$ is forward invariant. 

For any initial condition $x_0 \in \mathcal{V}_{\alpha_0}$, we  can now repeatedly apply \Cref{discrete_lyapunov_lemma} so that, for all $t\ge 0$,
\begin{align*}
    \overline{V}(x_t) &\leq \rho^t \overline{V}(x_0) + \sum_{k=0}^{t-1} \rho^k L \gamma \leq \rho^t \overline{V}(x_0) + \frac{L\gamma}{1-\rho}.
\end{align*}
\end{proof}
}
We conclude by noting that the discussion on practical stability  following
\Cref{continuous_lyapunov_proposition} also applies  to 
\Cref{discrete_lyapunov_proposition}.

% if one wants to show a result regarding the behavior $\|x_t\|$ directly, then a similar remark as \Cref{remark:lyap_convergence_of_xt}, imposing extra assumptions on the encoder, applies also to 
% \Cref{discrete_lyapunov_proposition}.

\section{Latent Barrier Functions}

% As in the Lyapunov case, our main results are conditions on the encoding and decoding maps $E$ and $D$, and latent dynamics $f_z$, that guarantee that safe control inputs for the latent system (i.e. controls which satisfy a control barrier function inequality) are also safe when applied to the original system $\dot{x}=f(x,u)$ or $x_{t+1}=f(x_t,u_t)$. We again assume that $f$ and $f_z$ share the same action space $\mathcal{U}\subseteq\mathbb{R}^{n_u}$.

We now move from the Lyapunov setting to that of barrier functions.
Here, we suppose there exists a safe set $\mathcal{S}_x \subseteq \R^{n_x}$ in the original space, and the objective is to verify the forward invariance of a non-empty subset of  $\mathcal{S}_x$ under the dynamics $f^{(\pi)}$.
From $\mathcal{S}_x$, we induce a latent safe set via the encoder in the following manner (cf.~\Cref{fig:latent_safe_set}):
\begin{align}
    \mathcal{S}_z := (E(\mathcal{S}_x^c))^c. \label{eq:latent_safe_set}
\end{align}
The latent safe set $\mathcal{S}_z$ has the property that $z \in \mathcal{S}_z$ if and only if there does \emph{not} exist any unsafe state $x \in \mathcal{S}_x^c$ such that $z = E(x)$.
This implies that $E^{-1}(\mathcal{S}_z) = \{ x \in \R^{n_x} \mid E(x) \in \mathcal{S}_z \} \subseteq \mathcal{S}_x$.\footnote{In general, $E^{-1}(\mathcal{S}_z)$ is an \emph{under-approximation} of $\mathcal{S}_x$.
Indeed, it is possible for $x \in \mathcal{S}_x$ but $x \not\in E^{-1}(\mathcal{S}_z)$, which happens if there exists $x \in \mathcal{S}_x$ and $x' \not\in \mathcal{S}_x$, but the encoder is unable to distinguish between them, i.e., $E(x) = E(x')$.}
Given a barrier function which certifies that the latent dynamics $f_z^{(\pi)}$ render (subsets of) $\mathcal{S}_z$ forward invariant, our goal is now to argue that system $f^{(\pi)}$ also renders (subsets of) $\mathcal{S}_x$ forward invariant. Again, we will divide our results between continuous and discrete time.

\subsection{Continuous Time Results for Barrier Functions}

We consider a continuously differentiable barrier function $h : \mathcal{D}_z \mapsto \R$ defined on an open set $\mathcal{D}_z \subseteq \R^{n_z}$ that contains the safe set $\mathcal{S}_z$.
The zero super-level set $\mathcal{C}_z := \{ z \in \mathcal{D}_z \mid h(z) \geq 0 \}$ of $h$ is assumed to be contained within the safe set $\mathcal{S}_z$, i.e.,   $\mathcal{C}_z \subseteq \mathcal{S}_z$, so that the zero super-level set of $h$ encodes a subset of the latent safe set $\mathcal{S}_z$.
As in \Cref{sec:lyapunov}, we carry over the definition $\mathcal{D}_x := E^{-1}(\mathcal{D}_z)$.
%
% Consider a continuous time system $\dot{x}=f(x,u)$ where $f:\mathbb{R}^{n_x}\times\mathbb{R}^{n_u}\rightarrow\mathbb{R}^{n_x}$, an encoding map $E:\mathbb{R}^{n_x}\rightarrow\mathbb{R}^{n_z}$, a decoding map $D:\mathbb{R}^{n_z}\rightarrow\mathbb{R}^{n_x}$, and a latent dynamics model $\dot{z}=f_z(z,u)$ where $f_z:\mathbb{R}^{n_z}\times\mathbb{R}^{n_u}\rightarrow\mathbb{R}^{n_z}$. We again assume that $n_z <n_x$. We are interested in the following class of control barrier functions.
%
The following definition of barrier functions is standard, see e.g., \cite{ames2019cbf}. 
\begin{df}[Continuous Time Barrier Function]\label{continuous_cbf_def}
Let $\alpha > 0$.
The function $h : \mathcal{D}_z \mapsto \R$ is a $\alpha$-barrier function if $\langle\nabla h(z),f^{(\pi)}_z(z)\rangle\geq-\alpha h(z)$ for all $z \in \mathcal{D}_z$.
% $\sup_{u\in\mathcal{U}}\dot{h}(z,u)=\sup_{u\in\mathcal{U}}\langle\nabla h(z),f_z(z,u)\rangle\geq-\alpha h(z)$ where $\alpha>0$, for all $z\in\mathcal{D}_z$. Furthermore, $h(x)\geq0$ on a `safe' subset $\mathcal{C}_z\subset\mathcal{D}_z$, and $h(z)<0$ on $\mathcal{D}_z\setminus\mathcal{C}_z$. If the domain of $h$ is in the output space of the encoder, we call $h$ a continuous time latent control barrier function.
\end{df}

Similar to the previous section, we can design a latent space controller $\pi(z)$ using the framework of control barrier functions~\cite{ames2019cbf}.
%, e.g., via the optimization problem $\pi(z) =\argmax_{u \in \mathcal{U}} \langle \nabla h(z), f_z(z, u) \rangle$. 
The next lemma characterizes the relationship between upper bounds on the conjugate dynamics error (cf.~\Cref{def:continuous_time_dynamics_error}) and its 
implication on the time derivative of the %\emph{almost} barrier 
function $\overline{h} = h \circ E$.
\begin{lmm}[Barrier Condition (Continuous Time)]\label{continuous_cbf_lemma} 
% Consider encoding map $E:\mathbb{R}^{n_x}\rightarrow\mathbb{R}^{n_z}$, original dynamics $f:\mathbb{R}^{n_x}\times\mathbb{R}^{n_u}\rightarrow\mathbb{R}^{n_x}$, latent dynamics $f_z:\mathbb{R}^{n_z}\times\mathbb{R}^{n_u}\rightarrow\mathbb{R}^{n_z}$. Let $h:\mathcal{D}_z\subset\mathbb{R}^{n_z}\rightarrow\mathbb{R}$ be an $L$-Lipschitz function that satisfies $\sup_{u\in\mathcal{U}}\langle\nabla h(z),f(z,u)\rangle\geq-\alpha h(z)$ where $\alpha>0$, for all $z\in\mathcal{D}_z$. With $\mathcal{D}_x=E^{-1}(\mathcal{D}_z)$, if either the conjugate dynamics error of the encoding or inverse encoding, \cref{encoder_form_error} or \cref{decoder_form_error}, are upper bounded by $\gamma$ under the latent feedback controller $\pi(z)\triangleq\arg\max_{u\in\mathcal{U}}\langle\nabla h(z), f_z(z,u)\rangle$, then
Let $\dot{x}=f^{(\pi)}(x)$ be continuous on $\mathcal{D}_x$ and $h : \mathcal{D}_z \mapsto \R$ be an $\alpha$-barrier function (cf.~\Cref{continuous_cbf_def}).
Suppose either that 
(i) $h$ is $L$-Lipschitz on $\mathcal{D}_z$ and $(f_z, \mathcal{D}_x)$ is $\gamma$-forward-conjugate (cf.~\Cref{encoder_form_error}),
or (ii) $h \circ E$ is $L$-Lipschitz on $\mathcal{D}_x$ and 
$(f_z, \mathcal{D}_x)$ is $\gamma$-backward-conjugate (cf.~\Cref{decoder_form_error}).
Then, for all $x \in \mathcal{D}_x$, % it holds that
\begin{equation}\label{continuous_cbf}
    \langle\nabla \overline{h}(x), f^{(\pi)}(x) \rangle+\alpha\overline{h}(x)\geq -L\gamma,
\end{equation} 
where the function $\overline{h} := h \circ E$ is defined over the domain $\mathcal{D}_x$, i.e., $\overline{h}:\mathcal{D}_x\mapsto \R$. 
%
% If \cref{decoder_form_error} is the tighter bound, assume that $h\circ E$ is $L$-Lipschitz.
\end{lmm}
\toggle{
\begin{proof}
% We first observe that
% \begin{align*}
%     \sup_{u \in \mathcal{U}} \langle \nabla \bar{h}(x), f(x, u) \rangle &\geq \langle \nabla \bar{h}(x), f(x, \pi(E(x))) \rangle \\
%     &= \langle \nabla \bar{h}(x), f^{(\pi)}(x) \rangle,
% \end{align*}
% where $\pi$ is defined in \eqref{eq:cbf_pi_continuous}.
% %
% The remainder of
Similar to \Cref{continuous_lyapunov_lemma}, and hence omitted.
% The proof is nearly identical to \Cref{continuous_lyapunov_lemma}, and hence we omit the details.
%
% We first show that \cref{encoder_form_error} implies \cref{continuous_cbf}.
% \begin{align*}
% \sup_{u\in\mathcal{U}}\dot{\overline{h}}(x,u)=\left\langle\frac{\partial E}{\partial x}(x)^T\nabla V(E(x)),f(x,u)\right\rangle\\
% =\left\langle\nabla V(E(x)), \frac{\partial E}{\partial x}(x)f(x,u)-f_z(E(x),u)\right\rangle\\
% +\bigg\langle\nabla V(E(x)),f_z(E(x), u)\bigg\rangle\\
% \geq-L\gamma-\alpha h(E(x))
% \end{align*}
% We next show that \cref{decoder_form_error} implies \cref{continuous_cbf}. By the previous steps,
% \begin{align*}
% \left\langle\frac{\partial E}{\partial x}(x)^T\nabla V(E(x)),f(E(x),u)-\frac{\partial E}{\partial x}(x)^\dagger f_z(E(x),u)\right\rangle\\
% +\bigg\langle\nabla V(E(x)), f_z(E(x),u)\bigg\rangle\geq -L\gamma-\alpha h(E(x))
% \end{align*}
\end{proof}
}

We use \Cref{continuous_cbf_lemma} to prove that sufficiently-low conjugate dynamics error means a latent barrier function yields a forward invariant safe-set for the original system.
%.as long as the encoding map is continuous and positive definite.
%
\begin{prp}[Safety Guarantee (Continuous Time)]
\label{continuous_cbf_proposition} 
% Consider latent dynamics $\dot{z}=f(z,u)$ and an $L$-Lipschitz latent control barrier function which satisfies \Cref{continuous_cbf_def}. Assume that the encoding map $E$ satisfies \Cref{positive_definite_encoder} on domain $\mathcal{D}_x=E^{-1}(\mathcal{D}_z)$. Also assume that \Cref{continuous_cbf_lemma} holds with conjugate dynamics error bounded by $\gamma$. Consider the candidate control barrier function $\overline{h}=h\circ E$. Then $\overline{h}$ certifies forward invariance of the set  
% \begin{equation}
%     \mathcal{C}_x=\{x\in\mathcal{D}_x:\overline{h}(x)\geq L\gamma/\alpha\}
% \end{equation} 
Let $\dot{x}=f^{(\pi)}(x)$ and $h:\mathcal{D}_z\mapsto \mathbb{R}$ satisfy the hypothesis of \Cref{continuous_cbf_lemma}, and suppose that  $\dot{x} = f^{(\pi)}(x)$ is forward complete.
%Furthermore, assume that $\mathcal{D}_z$ is compact. 
Suppose also that $\mathcal{C}_z$ is robustly included in $\mathcal{S}_z$ in the sense that $\mathcal{C}_z^\gamma := \{ z \in \mathcal{D}_z \mid h(z) \geq -L\gamma/\alpha \} \subseteq \mathcal{S}_z$. %and that the set $\mathcal{C}_z^\gamma$ is compact and non-empty.
%under bounded control input $\sup_{t\geq0}\|u_t\|_\infty<\infty$, and that the controller $\pi(z)$ is continuous over $\mathcal{D}_z$. 
Then, the set $\mathcal{C}_x^\gamma := \{ x \in \mathcal{D}_x \mid \overline{h}(x) \geq -L\gamma/\alpha\}$ is forward invariant.
\end{prp}
\toggle{
\begin{proof}
Similar to the proof of \Cref{continuous_lyapunov_proposition}, the comparison lemma gives us $\overline{h}(x_t)\ge (\overline{h}(x_0)+L\gamma/\rho)\exp(-\rho t)-L\gamma/\rho$ for all $t\ge 0$, i.e., $\mathcal{C}_x^\gamma$ is forward invariant.
%Define $h'(x) := \bar{h}(x) + L\gamma/\alpha$.
%By \Cref{continuous_cbf_lemma},
%we have 
%$\langle \nabla h'(x), f^{(\pi)}(x) \rangle \geq -\alpha h'(x)$.
%By the comparison lemma \cite[Theorem 3.3]{khalil2002nonlinear}, we have $h'(x(t)) \geq \exp(-\alpha t) h'(x(0))$.
%Therefore, if $x(0) \in \mathcal{C}_x^\gamma$, then $h'(x(0)) \geq 0 \Longrightarrow h'(x(t)) \geq 0$ for all $t$, from which the claim follows.
\end{proof}
}
The implication of \Cref{continuous_cbf_proposition}
is that the system is safe since $\mathcal{C}_x^\gamma \subseteq E^{-1}(\mathcal{S}_z) \subseteq \mathcal{S}_x$, i.e., a subset $\mathcal{C}_x^\gamma$ of the original safe-set $\mathcal{S}_x$ is rendered forward invariant.
%
% We finish the section with the following fact.
% \begin{prp}
% The set $\mathcal{C}_x^\gamma$ (cf.~\Cref{continuous_cbf_proposition})
% satisfies $\mathcal{C}_x^\gamma \subseteq \mathcal{S}_x$.
% \end{prp}
% \begin{proof}
% Fix $x \in \mathcal{C}_x^\gamma$ and suppose $x \not\in \mathcal{S}_x$. 
% % We have that $E(x) \in \mathcal{S}_z$
% Then, this means $E(x) \in \mathcal{S}_z^c$, which contradicts
% $x \in \mathcal{C}_x^\gamma \Longrightarrow E(x) \in \mathcal{S}_z$.
% \end{proof}

\subsection{Discrete Time Results for Barrier Functions}

% Consider a discrete time system $x_{t+1}=f(x_t,u_t)$, where $f:\mathbb{R}^{n_x}\times\mathbb{R}^{n_u}\rightarrow\mathbb{R}^{n_x}$. We consider an encoding map $E:\mathbb{R}^{n_x}\rightarrow\mathbb{R}^{n_z}$, a decoding map $D:\mathbb{R}^{n_z}\rightarrow\mathbb{R}^{n_x}$, and latent dynamics $z_{t+1}=f(z_t,u_t)$ where $f:\mathbb{R}^{n_z}\times\mathbb{R}^{n_u}\rightarrow\mathbb{R}^{n_z}$. Next, we define a class of discrete time control barrier functions.
We now focus on discrete-time barrier functions, 
adopting the definition from \cite{zeng2021mpc,freire2023cbf}.
\begin{df}[Discrete Time Barrier Function]\label{discrete_cbf_def}
Let $\alpha \in (0, 1]$. The function $h : \mathcal{D}_z \mapsto \R$ is a $\alpha$-discrete-barrier function if    $h(f_z^{(\pi)}(z))-h(z)\geq-\alpha h(z)$  for all $z \in \mathcal{D}_z$.
% Let $h:\mathcal{D}_z\subset\mathbb{R}^{n_z}\rightarrow\mathbb{R}$ be a function that satisfies $\sup_{u\in\mathcal{U}}\Delta h(x,u)=\sup_{u\in\mathcal{U}}h(f_z(z,u))-h(z)\geq-\alpha h(x)$ where $\alpha\in(0,1]$, for all $z\in\mathcal{D}_z$. Furthermore, $h(x)\geq0$ on a `safe' subset $\mathcal{C}_z\subset\mathcal{D}_z$, and $h(z)<0$ on $\mathcal{D}_z\setminus\mathcal{C}_z$. If the domain of $h$ in the output space of the encoder, we call $h$ a discrete time latent control barrier function.
\end{df}
%Analogous to the continuous time case \eqref{eq:cbf_pi_continuous},
%a discrete time controller can be selected as a control barrier function:
%\begin{align}
%    \pi(z) = \argmax_{u \in \mathcal{U}} h(f_z(z, u)).
%\end{align}

The next lemma characterizes the relationship between upper bounds on the conjugate dynamics error (cf.~\Cref{def:discrete_time_dynamics_error}) and its 
implication on the time derivative of the %\emph{almost} barrier 
function $\overline{h} = h \circ E$.

\begin{lmm}[Barrier Condition (Discrete Time)]\label{discrete_cbf_lemma}
Let $x_{t+1} = f^{(\pi)}(x_t)$  and $h : \mathcal{D}_z \mapsto \R$ be a $\alpha$-discrete-CBF (cf. \Cref{discrete_cbf_def}).
% Set $\mathcal{D}_x := E^{-1}(\mathcal{D}_z) = \{ x \in \R^{n_x} \mid \exists z \in \mathcal{D}_z \text{ s.t. } E(x) = z \}$.
Suppose either that 
(i) $h$ is $L$-Lipschitz and $(f_z, \mathcal{D}_x)$ is $\gamma$-forward-conjugate in discrete-time (cf.~\Cref{encoder_form_error_discrete}),
or (ii) $h \circ E$ is $L$-Lipschitz and 
$(f_z, \mathcal{D}_x)$ is $\gamma$-backward-conjugate in discrete-time (cf.~\Cref{decoder_form_error_discrete}).
Then, for all $x \in \mathcal{D}_x$, it holds that
\begin{equation}\label{discrete_time_cbf}
    \overline{h}(f^{(\pi)}(x)) - \overline{h}(x) \geq -\alpha \overline{h}(x) - L\gamma,
\end{equation}
where the function $\overline{h} := h \circ E$ is defined over the domain $\mathcal{D}_x$, i.e., $\overline{h}:\mathcal{D}_x\mapsto \R$. 
% Consider encoding map $E:\mathbb{R}^{n_x}\rightarrow\mathbb{R}^{n_z}$, original dynamics $f:\mathbb{R}^{n_x}\times\mathbb{R}^{n_u}\rightarrow\mathbb{R}$, latent dynamics $f_z:\mathbb{R}^{n_z}\times\mathbb{R}^{n_u}\rightarrow\mathbb{R}^{n_z}$. Let $h:\mathcal{D}_z\subset\mathbb{R}^{n_z}\rightarrow\mathbb{R}$ be an $L$-Lipschitz function that satisfies $\sup_{u\in\mathcal{U}}h(f_z(z,u))-h(z)\geq-\alpha h(z)$ where $\alpha\in(0,1]$, for all $z\in\mathcal{D}_z$. With $\mathcal{D}_x=E^{-1}(\mathcal{D}_z)$, if either the conjugate dynamics error of the encoding or inverse encoding, \cref{encoder_form_error_discrete} or \cref{decoder_form_error_discrete}, are upper bounded by $\gamma$ under the latent feedback controller $\pi(z)\triangleq\arg\max_{u\in\mathcal{U}}h(f_z(z,u))-h(z)$, then
% \begin{equation}\label{discrete_time_cbf}
%     \sup_{u\in\mathcal{U}}\overline{h}(f(x,u))-\overline{h}(x)\geq-\alpha h(x) -L\gamma
% \end{equation}
% for $\overline{h}=h\circ E$. If \cref{decoder_form_error_discrete} is the tighter bound, assume that $h\circ E$ is $L$-Lipschitz.
\end{lmm}
\toggle{
\begin{proof}
Similar to \Cref{discrete_lyapunov_lemma}, and hence omitted.
% The proof is nearly identical to \Cref{discrete_lyapunov_lemma}, and hence we omit the details.
% We first show that \cref{encoder_form_error_discrete} implies \cref{discrete_time_cbf}.
% \begin{align*}
% \sup_{u\in\mathcal{U}}\Delta\overline{h}(x,u)=\sup_{u\in\mathcal{U}}h(E(f(x,u)))-h(E(x))\\
% \geq-L\gamma-\sup_{u\in\mathcal{U}}h(f_z(E(x), u))-h(E(x))\\
% \geq-L\gamma-\alpha \overline{h}(x)
% \end{align*}
% Next, we show that \cref{decoder_form_error_discrete} implies \cref{discrete_time_cbf}
% \begin{align*}
% \sup_{u\in\mathcal{U}}\Delta\overline{h}(x,u)=\sup_{u\in\mathcal{U}}\overline{h}(f(x,u))-\overline{h}(x)\\
% =\sup_{u\in\mathcal{U}}\overline{h}(f(x,u))-h((E\circ E^{-1})(f_z(E(x),u)))\\
% +h(f_z(E(x),u))-h(E(x))\geq -L\gamma-\alpha\overline{h}(x)
% \end{align*}
\end{proof}
}

Finally, we mirror \Cref{continuous_cbf_proposition}
to construct a forward invariant safe-set in discrete time.

\begin{prp}[Safety Guarantee (Discrete Time)]\label{discrete_cbf_proposition} 
%
% Consider latent dynamics $\dot{z}=f(z,u)$ and an $L$-Lipschitz latent control barrier function which satisfies \Cref{discrete_cbf_def}. Assume that the encoding map satisfies \cref{positive_definite_encoder} on domain $\mathcal{D}x=E^{-1}(\mathcal{D}_z)$. Also assume that \cref{discrete_cbf_lemma} holds with conjugate dynamics error bounded by $\gamma$. Consider the candidate control barrier function $\overline{h}=h\circ E$. Then $\overline{h}$ certifies forward invariance of the set
% \begin{equation}
%     \mathcal{C}_x=\{x\in\mathcal{D}_x:\overline{h}(x)\geq L\gamma/\alpha\}
% \end{equation}
%
Let $x_{t+1} = f^{(\pi)}(x_t)$  and $h:\mathcal{D}_z\mapsto \mathbb{R}$ satisfy the hypothesis of \Cref{discrete_cbf_lemma}. Suppose 
that $\mathcal{C}_z$ is robustly included in $\mathcal{S}_z$ 
in the sense that 
$\mathcal{C}_z^\gamma := \{ z \mid h(z) \geq -L\gamma/\alpha \} \subseteq \mathcal{S}_z$ and that the set $\mathcal{C}_z^\gamma$ is  non-empty. Then, the set $\mathcal{C}_x^\gamma := \{ x \mid \overline{h}(x) \geq -L\gamma/\alpha\}$ is forward invariant.
\end{prp}
\toggle{
\begin{proof}
Similar to \Cref{discrete_lyapunov_proposition}, and hence omitted.
%Define $h'(x) := \bar{h}(x) + L\gamma/\alpha$.
%Applying \Cref{discrete_cbf_lemma}, we have that
%$h'(f^{(\pi)}(x)) \geq (1-\alpha) h'(x)$ for all $x \in \mathcal{D}_x$.
%Hence by induction if $x_0 \in \mathcal{C}_x^\gamma$, then $x_t \in \mathcal{C}_x^\gamma \,\, \forall t \in \mathbb{N}$.
% Consider the substitution $h'(x)=\overline{h}(x)-L\gamma/\alpha$. We therefore have $\sup_{u\in\mathcal{U}} h'(f(x,u))-L\gamma/\alpha-h'(x)+L\gamma/\alpha\geq-\alpha h'(x)+L\gamma-L\gamma$, which proves our result, since the zero-level set of $h'(x)$ is the $L\gamma/\alpha$-level set of $\overline{h}(x)$
\end{proof}
}

\section{Experiments}\label{sec:experiments}

We validate our approach on (1) cartpole stabilization via latent Lyapunov functions, and (2) a collision-avoidance task for a two vehicle system via latent barrier functions. 
% with our goal being to both verify \Crefrange{continuous_lyapunov_proposition}{discrete_cbf_proposition}, and to demonstrate that learning the encoding map, decoding map, and latent dynamics model using the forward-path and backwards-path losses (\Crefrange{encoder_form_error}{decoder_form_error_discrete}) leads to lower task-specific cost when $(E, D, f_z)$ are parameterized as neural networks. We verify the latter by performing an ablation study over multiple training runs.

% When training the encoding map, decoding map, and latent dynamics, it is not necessary to reconstruct all components of the system in the case that the control task only prescribes target behavior for only a subset of them.

% \subsection{Learning}
% To stabilize the cartpole in the following example, we learn $E$, $D$, and $f_z$. To make learning these models possible, we obtain a dataset of state-control trajectories from the original system (cartpole) under semi-random control inputs that are biased towards stabilization: we apply random control inputs and only add the resulting trajectories to our dataset if the pole stays upright for a fixed number of steps. For learning, we use empirical versions of \Crefrange{encoder_form_error}{decoder_form_error_discrete} that are averaged over the collected data, along with the following losses, also averaged over the dataset, that encourage $E$ and $D$ to be left and right inverses of eachother:
% \begin{equation}
%     \min_\theta \|(D_\theta\circ E_\theta)(x) - x\|^2 , \min_\theta \|(E\circ D)(z)-z\|^2
% \end{equation}
% where $z=E(x)$.

\subsection{Cartpole}
\label{sec:experiments:cartpole}

\toggletext{
% FULL VERSION VVV
The cartpole is a four-dimensional system with state $(x_1,\dot{x}_1,x_2,\dot{x}_2):=(x, \dot{x}, \theta, \dot{\theta})$ 
and dynamics
\begin{align}\label{eq:cartpole}
\ddot{\theta}&=\frac{g\sin(\theta)+cos(\theta)(-u-m_pl\dot{\theta}^2\sin(\theta)/(m_c+m_p))}{l(\frac{4}{3}-m_p\cos(\theta)^2/(m_c+m_p))},\nonumber\\
\ddot{x}&=\frac{u+m_pl(\dot{\theta}^2\sin(\theta)-\ddot{\theta}\cos(\theta))}{m_c+m_p},
\end{align}
where $g=9.8$, $m_c=0.5$, $m_p=0.05$, and $l=0.5$. The scalar $u$ is the control input which corresponds to the instantaneous force applied to the cart. 
%\textcolor{blue}{Though it is not necessary for this example, as we model the cartpole in discrete-time, we prove that the cartpole system is forward complete under bounded inputs in \cref{app:cartpole_forward_completeness}.}
We treat the cartpole as a discrete-time system by discretizing the continuous-time dynamics using a fourth-order Runge-Kutta method with timestep $0.02$.}
{
% CDC VERSION VVV
The cartpole is a four-dimensional system with state $(x_1,\dot{x}_1,x_2,\dot{x}_2):=(x, \dot{x}, \theta, \dot{\theta})$ and 
% \textcolor{blue}{(see full paper for details)}. 
input $u\in\R$ which denotes the instantaneous force applied to the cart. 
% \textcolor{blue}{We prove that the cartpole system is forward complete under bounded inputs, and discuss the compactness of $\mathcal{D}_z$, in the full paper.}
We treat the cartpole as a discrete-time system by discretizing the continuous-time dynamics using a fourth-order Runge-Kutta method with timestep $0.02$.}
We note that treating the cartpole as a 
continuous-time system can also be handled by
our results; in Appendix~\ref{app:cartpole_forward_completeness} we 
provide a self-contained proof showing
forward completeness of the cartpole
under bounded inputs.
\toggletext{
%
% FULL PAPER TEXT VVV
In this example we, illustrate the applicability of our main Lyapunov stability results
(cf.~\Cref{sec:lyapunov}) by (a) learning
a latent space and latent dynamics for the cartpole system, 
(b) using the latent dynamics to design a stabilizing controller via 
LQR and learning an associated Lyapunov function for the latent closed-loop system,
and (c) applying \Cref{discrete_lyapunov_proposition}
to estimate a valid invariant set and region of attraction for the original system. Encoding and decoding maps $E_\phi(x)$ and $D_\phi(z)$ are parameterized as two-layer multilayer perceptrons (MLPs), with encoder and decoder having a $64$-dimensional hidden state and $\tanh$ activations. The latent space, i.e., the output space of the encoding map and input space of the decoding map, is $2$-dimensional; setting $n_z=2$ is motivated in part
by the example from \cite[Sec 5.3]{tabuada2008approximate}
which analytically constructs an approximate $2$-dimensional dynamics for a similar type of cartpole system. The discrete-time latent dynamics $z^+=f_z^\phi(z,u)=A_\phi(z)z+B_\phi(z)u$ are also parameterized as a pair $(A_\phi(z),B_\phi(z))$ of two-layer MLPs with $2$-dimensional argument $z$ and $2\times 2$ matrix-valued output in the case of $A_\phi$, and $2\times 1$ vector-valued output in the case of $B_\phi$. Each has a $64$-dimensional hidden state, and $\tanh$ activations. 
Our estimated Lyapunov function is of the form 
$V_\phi(z)=F^2_\phi(z-z_\mathrm{eq})+0.1\|z-z_\mathrm{eq}\|^2$ using a $256$-dimensional, three-layer MLP $F_\phi(z):\mathbb{R}^{n_z}\mapsto\mathbb{R}$ with
$\tanh$ activations and no biases, so that $F_\phi(0) = 0$ by construction.}
{% CDC VERSION VVV
In this example, we illustrate the applicability of our main Lyapunov stability results
(cf.~\Cref{sec:lyapunov}) by 
(a) learning
a latent space and latent dynamics for the cartpole system, 
(b) using the latent dynamics to design a stabilizing controller via 
LQR and learning an associated Lyapunov function for the latent closed-loop system,
and (c) applying \Cref{discrete_lyapunov_proposition}
to estimate a valid invariant set and region of attraction for the original system. Encoding and decoding maps $E_\phi(x)$ and $D_\phi(z)$, latent dynamics $z^+=f_z^\phi(z,u)=A_\phi(z)z+B_\phi(z)u$, and Lyapunov function $V_\phi(z)=F^2_\phi(z-z_\mathrm{eq})+0.1\|z-z_\mathrm{eq}\|^2$ are all parameterized as neural networks.
We now briefly describe parts (a)-(c), with
the extended paper containing the full details.
}

% \label{sec:experiments:cartpole:learning_step}
\toggletext{
% FULL VERSION TEXT VVV
\paragraph{Learning latent state and dynamics}
\hypertarget{target:experiments:cartpole:learning_step}{For} learning the latent state and dynamics, we first obtain a dataset of state-control trajectories $\mathscr{D}_\mathrm{rand}=\big((x^{(i)}_t,u^{(i)}_t)_{t=1}^{T_1}\big)_{i=1}^{N_1}$ from a simulated cartpole system under semi-random control inputs that are biased towards stabilization.
In particular, for the initial state/action pair we grid over $\{x\in\mathbb{R}^4\mid\|x\|_\infty\leq 0.1\}$ and $\{u\in\mathbb{R}\mid |u| \leq 3\}$, so that every combination of $(x, u)$ in the aforementioned sets is included as an initial pair in $\mathscr{D}_\mathrm{rand}$. We then apply uniformly random control inputs in the range $[-3, 3]$ for the remaining $T_1-2$ timesteps, and only add the resulting trajectories to our dataset if the pole is upright ($|x_2|<\pi/2$) for all $T_1=16$ states in the trajectory. Alongside the semi-random control dataset, we collect a drift dataset, where every state in the semi-random control dataset is evolved 
for one-step with zero control input: $\mathscr{D}_\mathrm{drift}=\big((x^{(i)}_t,0)_{t=1}^{T_2=2}\big)_{i=1}^{N_2=N_1T_1}$. This is to ensure that a non-trivial drift term $A_\phi(z)$ is learned.}{
% CDC VERSION VVV
\textit{Learning Procedure and Latent Control Design:}
\hypertarget{target:experiments:cartpole:learning_step}{We} obtain a dataset of state-control trajectories $\mathscr{D}_\mathrm{rand}=\big((x^{(i)}_t,u^{(i)}_t)_{t=1}^{T_1}\big)_{i=1}^{N_1}$ from a simulated cartpole system under semi-random control inputs that are biased towards stabilization. We also collect a drift dataset, where every state in the semi-random control dataset is evolved 
for one-step with zero control input: $\mathscr{D}_\mathrm{drift}=\big((x^{(i)}_t,0)_{t=1}^{T_2=2}\big)_{i=1}^{N_2=N_1T_1}$. This ensures that a non-trivial drift term $A_\phi(z)$ is learned. \hypertarget{:experiments:cartpole:lyapunov}{A} stabilizing controller latent controller is obtained around the equilibrium point $z_\mathrm{eq} := E_\phi(0)$ by computing an LQR controller $\pi(z)=K(z-z_\mathrm{eq})$ from the matrices $A_\phi(z_\mathrm{eq}),B_\phi(z_\mathrm{eq})$.
We estimate a latent Lyapunov function 
% %V(z)=(z-z_\mathrm{eq})^\T(L(z)L(z)^\T+I_2 )(z-z_\mathrm{eq})$ 
% $V(z)=F(z-z_\mathrm{eq})^2+0.1\|z-z_\mathrm{eq}\|^2$ using a $256$-dimensional, three-layer, MLP $F(z):\mathbb{R}^{n_z}\rightarrow\mathbb{R}$ with no biases. 
from a dataset of latent trajectories rolled-out under the LQR controller.
}

\toggle{
We now describe the loss functions we use for learning.
We first minimize the weighted sum of empirical, multi-step versions of \Crefrange{encoder_form_error}{decoder_form_error_discrete} that are averaged over the dataset. Specifically, let $U_{j,k}=[u_{j}\ u_{j+1}\ \hdots\ u_{k-1}]\in\mathbb{R}^{n_u\times{k-j}}$ where $1\leq j < k \leq T$. Define $z_\phi^{t}:\mathbb{R}^{n_z}\times\mathbb{R}^{n_u\times (t-1)}\rightarrow\mathbb{R}^{n_x}$, where $z_\phi^t(z_1,U_{1,t})=f^\phi_z(f^\phi_z(\hdots f^\phi_z(z_1, u_{1}),\hdots), u_{t-2}), u_{t-1})$
evolves the latent dynamics model $f_z^\phi$ for $t \geq 1$ steps using the inputs $U_{1,t}$.
%
% Let the dataset $\mathscr{D} := \mathscr{D}_{\mathrm{rand}} \cup \mathscr{D}_{\mathrm{drift}}$.
Consider the following dynamics losses
defined for a given dataset $\mathscr{D} = ((x_t^{(i)}, u_t^{(i)})_{t=1}^{T_i})_{i=1}^{N}$:
%\sum_{\underset{\in\mathscr{D}}{(x_t, u_t)_{t=1}^T}}
\begin{align*}
&L_\mathrm{fwd}\!\!\:(\phi; \!\mathscr{D})\!=\!\!\sum_{i=1}^N\!\sum_{t=1}^{T_i\!-\!1}\!\sum_{k=1}^{T_i\!-\!\!\:t}\frac{\!\big\|\!\!\;E_\phi(x^{(i)}_{k+t})\!-\!z_\phi^t\big(E_\phi(x^{(i)}_k),\!\!\;U^{(i)}_{k,k+t}\big)\!\!\:\big\|^2}{N},\\
&\begin{aligned}L_\mathrm{bwd}\!\!\:(\phi; \!\mathscr{D})\!=\!\!\sum_{i=1}^{N}\!\sum_{t=1}^{T_i\!-\!1}\!\sum_{k=1}^{T_i\!-\!\!\:t}\frac{\!\big\|x^{(i)}_{k+t}\!-\!D_\phi\!\!\:\big(z_\phi^t\!\!\:\big(\!\!\;E_\phi\!\!\;(x^{(i)}_k),\!\!\;U^{(i)}_{k,k+t}\big)\!\big)\!\!\:\big\|^2}{N}.
\end{aligned}
\end{align*}
%where $\vec{u}$ denotes a sequence of control inputs to be applied to the dynamics. 
Along with the dynamics losses $L_\mathrm{fwd}$ and $L_\mathrm{bwd}$, which minimize the forwards and backwards conjugacy errors, we also encourage $E_\phi$ and $D_\phi$ to be left and right inverses:
\begin{align*}
    L_\mathrm{left}(\phi; \mathscr{D})&=\sum_{i=1}^N \sum_{t=1}^{T_i}\frac{\big\|(D_\phi\circ E_\phi)(x^{(i)}_t)-x^{(i)}_t\big\|^2}{NT_i},\\
    L_\mathrm{right}(\phi; \mathscr{D})&=\sum_{i=1}^N\sum_{t=1}^{T_i}\frac{\big\|(E_\phi\circ D_\phi)(E_\phi(x^{(i)}_t))-E_\phi(x^{(i)}_t)\big\|^2}{NT_i}.
    %\mathbb{E}_{x\in\mathscr{D}}\|(D\circ E)(x) - x\|^2 , \quad \mathbb{E}_{z\in E(\mathscr{D})}\|(E \circ D)(z)-z\|^2.
\end{align*}
In order to control the geometry of the latent space, we add losses that encourage the equilibrium to be approximately the origin in the latent space, and that encourage the encoded dataset to have approximately identity covariance:
\begin{align*}
L_\mathrm{ori}(\phi) &= \left\|E_\phi(0)\right\|, \\
L_\mathrm{iso}(\phi; \mathscr{D}) &=\left\|I_{n_z\times n_z}-\sum_{i=1}^N\sum_{t=1}^{T_i} \frac{E_\phi(x_t^{(i)})E_\phi(x_t^{(i)})^\T}{NT_i} \right\|.
\end{align*}
We learn the models $E_\phi(x), D_\phi(z), f_z^\phi(z, u)$ via a weighted sum of the aforementioned losses, with  
dataset $\mathscr{D}'$ defined as $\mathscr{D}' := \mathscr{D}_{\mathrm{rand}} \cup \mathscr{D}_{\mathrm{drift}}$:
\begin{align*}
L(\phi)=&\ \lambda_1 L_{\mathrm{fwd}}(\phi; \mathscr{D}')+ \lambda_2 L_\mathrm{bwd}(\phi; \mathscr{D}')+\\
&\ \lambda_3 L_\mathrm{left}(\phi; \mathscr{D}_\mathrm{rand})+\lambda_4 L_\mathrm{right}(\phi; \mathscr{D}_\mathrm{rand})+\\
&\ \lambda_5 L_\mathrm{ori}(\phi)+\lambda_6 L_\mathrm{iso}(\phi; \mathscr{D}_\mathrm{rand}).
\end{align*}
For our cartpole experiment, since we intend to use the forward conjugacy error when applying \cref{discrete_lyapunov_proposition}, we set $\lambda_1=5$, and $\lambda_i=1$ for $i \in \{2, \dots, 6\}$.
}

\toggle{\paragraph{Latent control design and Lyapunov analysis}}
\toggletext{
% FULL VERSION TEXT VVV
\hypertarget{target:experiments:cartpole:lyapunov}{To} obtain a stabilizing latent controller around the equilibrium point $z_\mathrm{eq} := E_\phi(0)$, we compute an LQR controller $\pi(z)=K(z-z_\mathrm{eq})$ by solving the discrete-time infinite-horizon Riccati equation with dynamics matrices $A_\phi(z_\mathrm{eq}),B_\phi(z_\mathrm{eq})$ and costs $Q=I_{n_x}$, $R=I_{n_u}$.
We estimate a latent Lyapunov function 
% %V(z)=(z-z_\mathrm{eq})^\T(L(z)L(z)^\T+I_2 )(z-z_\mathrm{eq})$ 
% $V(z)=F(z-z_\mathrm{eq})^2+0.1\|z-z_\mathrm{eq}\|^2$ using a $256$-dimensional, three-layer, MLP $F(z):\mathbb{R}^{n_z}\rightarrow\mathbb{R}$ with no biases. 
$V_\phi(z)$ by minimizing $L_{\mathrm{Lyap}}(\phi) = \sum_{z_t^{(i)}} \max(0, V_\phi(f_z^{(\pi)}(z_t^{(i)}))-\rho V_\phi(z_t^{(i)}))^2$ over a dataset of latent trajectories rolled-out under the LQR controller $\pi$, from initial conditions densely sampled from the set $\{z\in\mathbb{R}^{n_z}\mid\|z-z_\mathrm{eq}\|_\infty\leq 1.5\}$. 
We define
$\mathcal{D}_z$, the corresponding domain of the Lyapunov function $V_\phi$, as the convex hull of the union of these latent trajectories; more details in Appendix~\ref{Dx_and_Dz_details}. We note that $V_\phi$ is learned with the latent dynamics model obtained in Step \hyperlink{target:experiments:cartpole:learning_step}{(a)} held fixed, i.e., we do not jointly train $f_z^{\phi}$ and $V_\phi$.
%\textcolor{red}{What loss do we use to learn $V$?}
% I'll add this, just the dynamics loss, since it's already positive definite
}{\color{blue}
% CDC VERSION VVV
\color{black}}

% FULL PAPER TEXT VVV
\toggletext{
\paragraph{Application of \Cref{discrete_lyapunov_proposition}}
Our main goal is to apply \Cref{discrete_lyapunov_proposition} to reason about the stability properties of our LQR controller computed from the learned latent dynamics.
To do this, there are two main pre-conditions:
(i) estimate the forward-conjugacy error (cf.~\Cref{def:discrete_time_dynamics_error})
and (ii) verify that the learned Lyapunov function $V$
is a valid Lyapunov function (cf.~\Cref{discrete_time_lyapunov_def}) for the latent dynamics.\footnote{Moving forward, we drop the $\phi$-subscript to reduce notational clutter, noting that the models $V$, $E$, and $f_z$ come from learning procedure described in Steps \hyperlink{target:experiments:cartpole:learning_step}{(a)} and \hyperlink{target:experiments:cartpole:lyapunov}{(b)}.} For simplicity, we choose to use 
numerical gridding in order to verify (i) and (ii).
The key challenge that arises with numerical gridding is that 
both the sets $\mathcal{D}_z$ and $\mathcal{D}_x$ can be unbounded sets, and hence we must restrict our attention to bounded subsets. To do this, we roll out trajectories starting from bounded initial conditions, and estimate approximate invariant sets containing these trajectories; see Appendix~\ref{Dx_and_Dz_details} for more details.\footnote{Due to the approximate nature of the computation of these sets, we do not technically rigorously verify the stability of the cartpole system under our latent controller; instead, this demonstration serves to illustrate 
how our theorem would be applied if its pre-conditions were formally verified.}
%
% With the disclaimer in place, 
%
Using our estimates of $\mathcal{D}_z$ and $\mathcal{D}_x$,
we identify a forward invariant region defined by $\overline{V}(x)\leq\alpha_0$ where $\alpha_0:=\sup\{\alpha>0\mid\mathcal{V}_\alpha\subseteq\mathcal{D}_x\}=6.71$, i.e., the largest level set contained in $\mathcal{D}_x$.
\Cref{discrete_lyapunov_proposition} also identifies an over-approximation of the smallest attractive set contained within $\overline{V}(x)\leq\alpha_0$, namely $\overline{V}(x)\leq L\gamma/(1-\rho)$. 
We estimate the constant
$L$ (the Lipschitz constant of $V$) to be $25.60$,
$\gamma$ (the forward-conjugate error) as $0.0183$, and $\rho$ (the decay parameter of the latent Lyapunov function) as $0.85$. 
Furthermore, we also visualize a possible source of conservativeness in our bound.
In particular, recall in the proof of \Cref{discrete_lyapunov_lemma} that we write $\Delta\overline{V}(x)=\Delta V(E(x))+R(x)$, where $R(x)=V(E(f^{(\pi)}(x))-V(f^{(\pi)}_z(E(x)))$, and bound $\max_{x\in\mathcal{D}_x}|R(x)|/(1-\rho)$ with $L\gamma/(1-\rho)=3.13$. 
To quantify the conservativeness of this bound,
we compare our bound to direct computation of $\max_{x\in\mathcal{D}_x}|R(x)|/(1-\rho)=1.93$ over trajectories in \Cref{fig_lyapunov_convergence}. We observe that for small values of $(1-\rho)$ ($0.15$ in this case), bounding $|R(x)|$ with $L\gamma$ yields a reasonably tight over-approximation when compared to direct computation.

Additionally, to assess the convergent behavior of the original system guaranteed by our theory, we plot the zero set $E^{-1}(0)$ (cf.\ the discussion at the end of \Cref{sec:lyapunov:continuous}), which yields the manifold the system can move in without affecting $\overline{V}$, and therefore the manifold to which the cartpole system converges.
Notably, we find that system trajectories are convergent for a significantly larger set than what our theorem can identify, and also that those trajectories converge to a significantly smaller set than what our theory identifies, suggesting that further theoretical gains can be made on both ends. As well, system trajectories are observed to converge to the origin, even while in the zero-set $E^{-1}(0)$, as shown in \Cref{fig_lyapunov_convergence}, which is not captured by our theory. We leave further investigation of this phenomenon to future work.

Finally, we observe that the geometry of latent space trajectories dictates which $\rho$ are feasible to obtain a Lyapunov function for, and is therefore critical to obtaining a tight over-approximation of the attractive region, given by $L\gamma/(1-\rho)$. It is also important to train a well-behaved encoder and latent model over as large a $\mathcal{D}_x$ as possible, to ensure that the obtained forward invariant and attractive regions are non-trivial. We find this latter condition to be challenging to balance with the fidelity of the latent model, for the cartpole system. Understanding the limitations of latent representations, and to what extent their geometry can be shaped, may be interesting directions for future work.}{
% CDC PAPER TEXT VVV
\textit{Application of \Cref{discrete_lyapunov_proposition}:}
We apply \Cref{discrete_lyapunov_proposition} to reason about the full system in closed-loop with the latent LQR controller. This requires two main pre-conditions:
(i) estimating the forward-conjugacy error (cf.~\Cref{def:discrete_time_dynamics_error})
and (ii) verifying the validity of the learned Lyapunov function $V$ (cf.~\Cref{discrete_time_lyapunov_def}) for the latent dynamics. For simplicity, we use 
numerical gridding over trajectories in order to verify (i) and (ii). Because $\mathcal{D}_x$ may be an unbounded set, we restrict our attention to a bounded, invariant subset. % See the full paper for more details.\par

We identify a forward invariant region defined by $\overline{V}(x)\leq\alpha_0$ where $\alpha_0:=\sup\{\alpha>0\mid\mathcal{V}_\alpha\subseteq\mathcal{D}_x\}=6.71$, i.e., the largest level set contained in $\mathcal{D}_x$.
\Cref{discrete_lyapunov_proposition} also identifies an over-approximation of the smallest attractive set contained within $\overline{V}(x)\leq\alpha_0$, namely $\overline{V}(x)\leq L\gamma/(1-\rho)$. 
We estimate the constant $L$ (the Lipschitz constant of $V$) to be $25.60$, $\gamma$ (the forward-conjugate error) as $0.0183$, and $\rho$ (the decay parameter of the latent Lyapunov function) as $0.85$. 
We also visualize a source of conservativeness in our bound. 
In the proof of \Cref{discrete_lyapunov_lemma} (given in the full paper), we write $\Delta\overline{V}(x)=\Delta V(E(x))+R(x)$, where $R(x)=V(E(f^{(\pi)}(x))-V(f^{(\pi)}_z(E(x)))$, and bound $\max_{x\in\mathcal{D}_x}|R(x)|/(1-\rho)$ with $L\gamma/(1-\rho)=3.13$. 
To quantify the conservativeness of this bound,
we compare our bound to direct computation of $\max_{x\in\mathcal{D}_x}|R(x)|/(1-\rho)=1.93$ over trajectories in \Cref{fig_lyapunov_convergence}. We observe that for $1-\rho=0.15$, bounding $|R(x)|$ with $L\gamma$ yields a reasonably tight over-approximation when compared to direct computation.

Additionally, we plot the zero set $E^{-1}(0)$ (cf.\ the discussion at the end of \Cref{sec:lyapunov:continuous}), which yields the manifold the system can move in without affecting $\overline{V}$. System trajectories are observed to converge to the origin, even while in the zero-set $E^{-1}(0)$, as shown in \Cref{fig_lyapunov_convergence}, which is not captured by our theory. We leave further investigation of this phenomenon to future work.}

\toggletext{
% FULL TEXT VERSION VVV
\begin{figure}{}
\centering
\includegraphics[width=8.8cm]{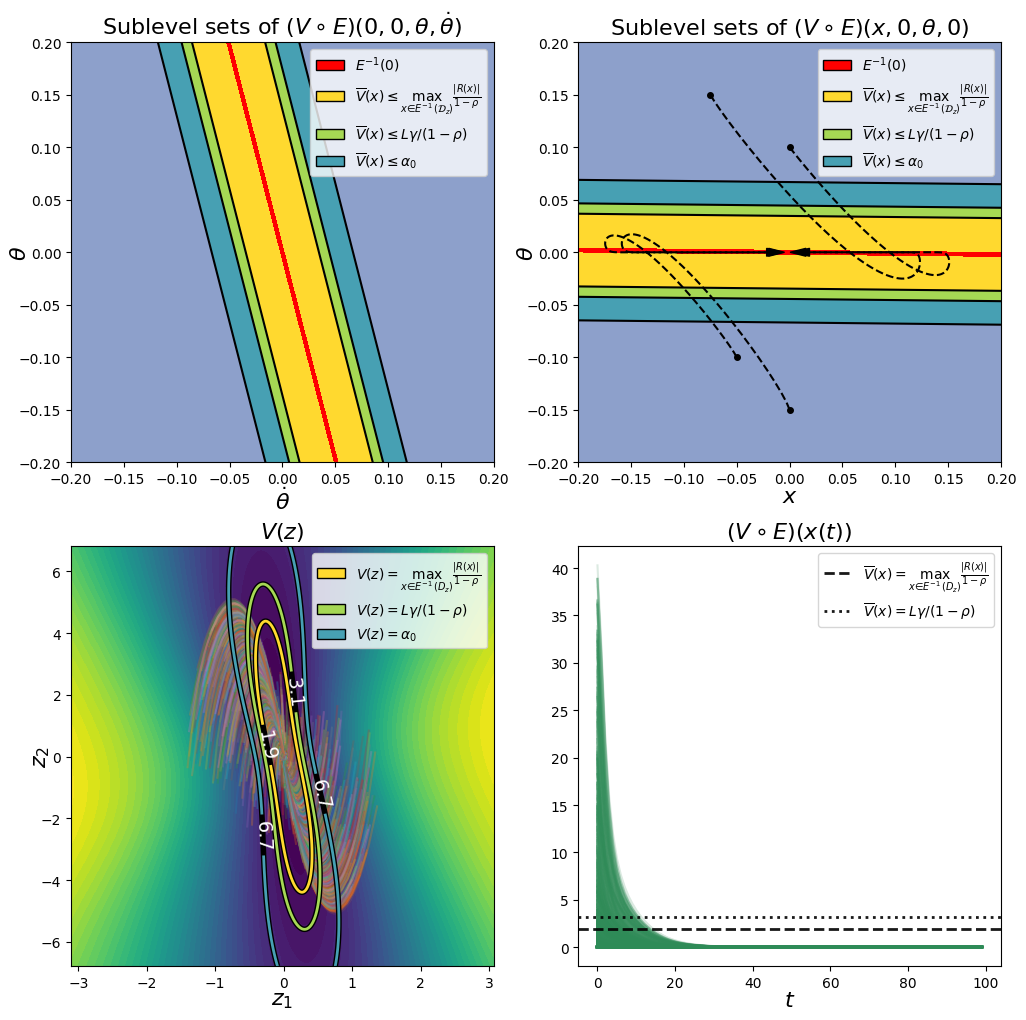}
%\vspace*{-7mm}
\caption{\emph{Top left:} Sub-level sets of $\overline{V}=V\circ E$ for the $(0,0,\theta, \dot{\theta})$-slice of the cartpole state space under LQR controller. Over-approximation of the attractive set, $\overline{V}(x)\leq L\gamma/(1-\rho)$, is compared to directly-computed bound on attractive set, $\max_{x\in \mathcal{D}_x}|R(x)|/(1-\rho)$. For $(1-\rho)=0.15$, bounding $|R(x)|$ with $L\gamma$ yields a reasonable over-approximation. Note the zero set $E^{-1}(0)$ in red; the system can move in this direction while keeping $\overline{V}$ constant. \emph{Top right:} Same as top-left, but for $(x,0,\theta,0)$-slice. Included system trajectories demonstrate convergent behavior: these trajectories converge to the origin even once entering the zero set $E^{-1}(0)$. \emph{Bottom left:} Trajectories $x_t$ projected into latent space, and level sets of $V$. \emph{Bottom right:} System trajectories converge to the zero-level set of $\overline{V}=V\circ E$; dashed line corresponds to directly-computed bound on the minimal attractive level set, dotted line corresponds to over-approximation $L\gamma/(1-\rho)$.
}
%\caption{\emph{Top left:} Vector field of $(0,0,x_2, \dot{x_2})$-slice of the cartpole state space under the LQR controller, with $V\circ E$ contour plotted under. The system is stable to a central line in the pendulum state. \emph{Top right:} Vector field of $(x_1,0,x_2,0)$-slice of the cartpole state space, also under the LQR controller. The system is stable to a line in position-angle space. \emph{Bottom left:} The latent Lyapunov function learned from latent LQR trajectories. \emph{Bottom right:} The state trajectories of the original system converge to the zero-level set of $V\circ E$. The grey-dashed line denotes the value under which trajectories do not satisfy the Lyapunov inequality, which corresponds conceptually to the $L\gamma/(1-\rho)$ level-set in our theorems.
%}
\label{fig_lyapunov_convergence}
%\vspace*{-5mm}
\end{figure}}{
% CDC VERSION VVV
\begin{figure}{}
\centering
\includegraphics[width=7.5cm]{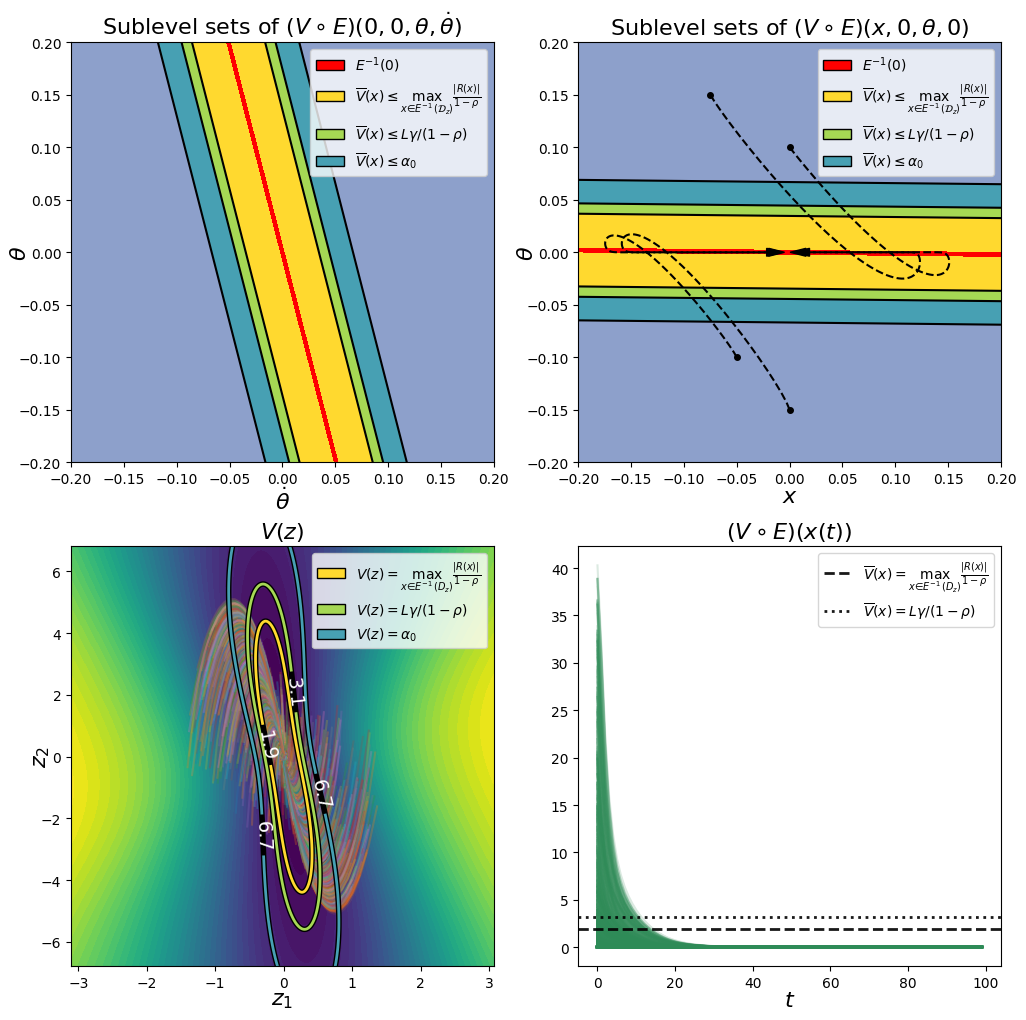}
%\vspace*{-7mm}
\caption{\small{\emph{Top left:} Sub-level sets of $\overline{V}=V\circ E$ for the $(0,0,\theta, \dot{\theta})$-slice of the cartpole state space under LQR controller. Over-approximation of the attractive set, $\overline{V}(x)\leq L\gamma/(1-\rho)$, is compared to directly-computed bound on attractive set, $\max_{x\in \mathcal{D}_x}|R(x)|/(1-\rho)$. For $(1-\rho)=0.15$, bounding $|R(x)|$ with $L\gamma$ yields a reasonable over-approximation. Note the zero set $E^{-1}(0)$ in red; the system can move in this direction while keeping $\overline{V}$ constant. \emph{Top right:} Same as top-left, but for $(x,0,\theta,0)$-slice. Included system trajectories demonstrate convergent behavior: these trajectories converge to the origin even once entering the zero set $E^{-1}(0)$. \emph{Bottom left:} Trajectories $x_t$ projected into latent space, and level sets of $V$. \emph{Bottom right:} System trajectories converge to the zero-level set of $\overline{V}=V\circ E$; dashed line corresponds to directly-computed bound on the minimal attractive level set, dotted line corresponds to over-approximation $L\gamma/(1-\rho)$.}
}
%\caption{\emph{Top left:} Vector field of $(0,0,x_2, \dot{x_2})$-slice of the cartpole state space under the LQR controller, with $V\circ E$ contour plotted under. The system is stable to a central line in the pendulum state. \emph{Top right:} Vector field of $(x_1,0,x_2,0)$-slice of the cartpole state space, also under the LQR controller. The system is stable to a line in position-angle space. \emph{Bottom left:} The latent Lyapunov function learned from latent LQR trajectories. \emph{Bottom right:} The state trajectories of the original system converge to the zero-level set of $V\circ E$. The grey-dashed line denotes the value under which trajectories do not satisfy the Lyapunov inequality, which corresponds conceptually to the $L\gamma/(1-\rho)$ level-set in our theorems.
%}
\label{fig_lyapunov_convergence}
%\vspace*{-5mm}
\end{figure}}

\subsection{Omni-directional Vehicle Avoidance}\label{sec:omni-directional_vehicle}
\toggletext{
% FULL VERISON TEXT VVV
Our next experiment consists of two planar vehicles, 
one which is an actively controlled omnidirectional vehicle~\cite{liu2008omni}.
The complete system state is described by a six-dimensional vector $x = (q_1, q_2) \in \R^6$, 
with the active car $q_1 = (p_1, \theta_1)$
and passive car $q_2 = (p_2,\theta_2)$, where $p_i \in \R^2$.
%
% omnidirectional vehicles \cite{liu2008omni}, one of which is actively controlled. The system state is described by a six-dimensional vector $(q_1,q_2)\in\mathbb{R}^6$, with active car $q_1\coloneqq(x_1,y_1,\theta_1)$ and passive car $q_2\coloneqq(x_2,y_2,\theta_2)$. 
The dynamics for the active car is $\dot{q}_1= G(\theta_1)B^{-\T}u + d(x)$, where
% \textcolor{red}{we can remove this formula for $G$ if we need space}
\begin{align*}
    G(\theta_1)&= \left[\begin{array}{ccc}
    \cos (\theta_1) & -\sin(\theta_1) & 0 \\
    \sin (\theta_1) & \cos(\theta_1) & 0 \\
    0 & 0 & 1
    \end{array}\right],
    % \\
    % B&=\left[\begin{array}{ccc}
    % 0 & r \cos (\pi / 6) & -r \cos (\pi / 6) \\
    % -r & r \sin (\pi / 6) & r \sin (\pi / 6) \\
    % \ell r & \ell r & \ell r
    % \end{array}\right].
\end{align*}
and $B$ is an invertible $3 \times 3$ 
matrix that depends only on
the radius of the robot's body ($\ell := 0.2$)
and radius of the each wheel ($r:=0.02$);
see the extended paper for the exact formula.
%
% the robot's geometry with $\ell :=0.2$ and $r:=0.02$ denoting the radius of the robot's body and the radius of each wheel, respectively. 
The disturbance $d(x)$ represents unmodeled dynamics
satisfying $\| d(x) \| \leq B_d$ for all $x \in \R^6$.
The control input $u\in [-B_u, B_u]^3$ denotes the angular velocity of each wheel.
The passive vehicle maintains a constant velocity:
$\dot{p}_2 = (\cos(\theta_2), \sin(\theta_2))$ and
$\dot{\theta}_2 = 0$.
We prove that the omni-directional vehicle system is forward complete under bounded control inputs in Appendix~\ref{app:omni_forward_completeness}.

In this example, instead of learning the encoder $(E, D)$ pair and latent dynamics $f_z$, we utilize the structure of the problem and fix the encoder to be 
$E(q) := (p_1 - p_2, \theta_1)$,
and the latent dynamics to be
\begin{align*}
    \dot{z} = f_z(z, u; \theta_2)=G(z_3)B^{-\T}u - (\cos\theta_2, \sin\theta_2, 0).
\end{align*}
Here, we note that $f_z$ is parameterized by $\theta_2$, which is due to the constant velocity assumption of the passive car. 

Our notion of safety in this problem is that the two cars keep a minimum distance $\beta>0$ away from each other, which yields the safe sets: 
\begin{align*}
    \mathcal{S}_x&=\left\{x \in \mathbb{R}^6 \mid\left\|p_1 - p_2 \right\| \geq \beta\right\},\\
    \mathcal{S}_z&= (E(\mathcal{S}_x^c))^c = \left\{z \in \mathbb{R}^3 \mid\left\|\left(z_1, z_2\right)\right\| \geq\beta\right\}.
\end{align*}
We design a latent CBF $\phi_{\beta'}(z)$ which operates on $z \in \mathbb{R}^3$ to render a subset $\mathcal{C}_z\subseteq\mathcal{S}_z$ forward invariant under $f_z$
via the following latent CBF quadratic program (CBF-QP):
%
% With a valid $\phi_{\beta'}(z)$, we can implement a safety filter by solving the latent CBF quadratic program (CBF-QP):
\begin{align*}
\label{eq:latent_cbf_ineq}
\pi_\beta(z; \theta_2) =\;
& \underset{u\in\mathcal{U}}{\mathrm{arg\,min}} 
& & |u - u_{\text{nom}}|^2   \notag\\
& \mathrm{s.t.}
& &\langle \nabla \phi_\beta(z), f_z(z, u; \theta_2) \rangle + \alpha \phi_\beta(z) \geq 0,
\end{align*}
where we assume $u_\text{nom}$ is generated by a nominal controller that achieves task-specific goals regardless of safety (e.g., proportional control that steers the car towards a goal). 
% We set $\alpha=1$ for the following experiments.

We propose the following candidate latent CBF:
\begin{align*}
    \phi_{\beta'}(z)=\left\|\left(z_1, z_2\right)\right\|-{\beta'}, \quad \mathcal{D}_z = \{z \in \R^3 \mid (z_1, z_2) \neq 0 \}.
\end{align*}
% Define the set $\mathcal{D}_z := \R^3 \setminus \{0\}$.
We prove in \toggletext{Appendix~\ref{sec:appendix:valid_CBF_proof}}{the full version} that if the input bound
$B_u \geq r \sqrt{3(1+\ell^2)} (1 + \alpha \beta')$,
then for all $z \in \mathcal{D}_z$,
$\sup_{u \in \mathcal{U}} \langle \nabla \phi_{\beta'}(z), f_z(z, u; \theta_2) \rangle + \alpha \phi_{\beta'}(z) \geq 0$.
Hence, the safety set $\mathcal{C}_z = \{ z \in \mathcal{D}_z \mid \phi_{\beta'}(z) \geq 0 \}$ is forward invariant under 
$f_z^{(\pi_{\beta'})}(z; \theta_2) := f_z(z, \pi_{\beta'}(z); \theta_2)$.
Next, it is straightforward to show
that the 
forward conjugacy error (cf.~\Cref{def:continuous_time_dynamics_error}) 
of our setup is bounded by $\gamma = B_d$. Additionally, $\phi_{\beta'}$ is $1$-Lipschitz. Hence, 
if
$\mathcal{C}_z^\gamma := \{ z \in \mathcal{D}_z \mid \phi_{\beta'}(z) \geq -B_d/\alpha \} \subseteq \mathcal{S}_z$, then by 
\Cref{continuous_cbf_proposition} the set
$\mathcal{C}_x^\gamma = \{ x \in \R^6 \mid \| p_1 - p_2 \| \geq \beta' - B_d/\alpha, \,\, x_6 = \theta_2 \}$ is forward invariant under the dynamics $f^{(\pi_{\beta'})}$.
For $\mathcal{C}_z^\gamma \subseteq \mathcal{S}_z$ to occur, we simply take $\beta' \geq \beta + B_d/\alpha$.

We illustrate this example in \Cref{fig:2car_avoidance}. 
Numerically, we choose $\alpha=2$, $B_d=1$, $\beta = 4.5$, $\beta' = 5$, and $B_u=5$. The disturbance is chosen adversarially as $d(x) = - B_d (p_1 - p_2)/\| p_1 - p_2 \|$,
which encourages the relative position between the cars to 
decrease.
The nominal control $u_{\mathrm{nom}}$
is a simple proportional controller to encourage the active car
to move towards the initial position of the passive car;
we arrange the initial condition so that 
using only $u_{\mathrm{nom}}$ for control will result in a safety violation.
We see in \Cref{fig:2car_avoidance} that the behavior predicted
by our theory is precise; by designing $\beta'$ to provide a robust safety margin as required by \Cref{continuous_cbf_proposition}, the active car is able to reach its target goal without exiting the safe set $\mathcal{S}_x$.}
{
% CDC VERSION VVV
Our next experiment consists of two planar vehicles, 
one which is an actively controlled omnidirectional vehicle~\cite{liu2008omni}.
The complete system state is described by a six-dimensional vector $x = (q_1, q_2) \in \R^6$, 
with the active car $q_1 = (p_1, \theta_1)$
and passive car $q_2 = (p_2,\theta_2)$, where $p_i \in \R^2$.
%
% omnidirectional vehicles \cite{liu2008omni}, one of which is actively controlled. The system state is described by a six-dimensional vector $(q_1,q_2)\in\mathbb{R}^6$, with active car $q_1\coloneqq(x_1,y_1,\theta_1)$ and passive car $q_2\coloneqq(x_2,y_2,\theta_2)$. 
The dynamics for the active car is $\dot{q}_1= G(\theta_1)B^{-\T}u + d(x)$ where $G(\theta_1)$ is an
orthogonal matrix,
% \textcolor{red}{we can remove this formula for $G$ if we need space}
and $B$ is an invertible $3 \times 3$ 
matrix that depends only on
the radius of the robot's body ($\ell := 0.2$)
and radius of the each wheel ($r:=0.02$);
see the extended paper for the exact expressions.
%
% the robot's geometry with $\ell :=0.2$ and $r:=0.02$ denoting the radius of the robot's body and the radius of each wheel, respectively. 
The disturbance $d(x)$ represents unmodeled dynamics
satisfying $\| d(x) \| \leq B_d$ for all $x \in \R^6$.
The control input $u\in [-B_u, B_u]^3$ denotes the angular velocity of each wheel.
The passive vehicle maintains a constant velocity:
$\dot{p}_2 = (\cos(\theta_2), \sin(\theta_2))$ and
$\dot{\theta}_2 = 0$. 
% \textcolor{blue}{In the full paper, we prove that the omni-directional vehicle system is forward complete under bounded control inputs, that $\pi(z)$ is bounded despite $\mathcal{D}_z$ lacking compactness, and that the CBF controller is continuous.}

In this example, instead of learning the encoder $(E, D)$ pair and latent dynamics $f_z$, we utilize the structure of the problem and fix the encoder to be 
$E(q) := (p_1 - p_2, \theta_1)$,
and the latent dynamics to be $\dot{z} = f_z(z, u; \theta_2)=G(z_3)B^{-\T}u - (\cos\theta_2, \sin\theta_2, 0)$.
%\begin{align*}
%    \dot{z} = f_z(z, u; \theta_2)=G(z_3)B^{-\T}u - (\cos\theta_2, \sin\theta_2, 0).
%\end{align*}
Here, we note that $f_z$ is parameterized by $\theta_2$, which is due to the constant velocity assumption of the passive car. Our notion of safety in this problem is that the two cars keep a minimum distance $\beta>0$ away from each other, which yields the safe sets: $\mathcal{S}_x=\left\{x \in \mathbb{R}^6 \mid\left\|p_1 - p_2 \right\| \geq \beta\right\}$ and $\mathcal{S}_z=(E(\mathcal{S}_x^c))^c = \left\{z \in \mathbb{R}^3 \mid\left\|\left(z_1, z_2\right)\right\| \geq\beta\right\}$.
%\begin{align*}
%    \mathcal{S}_x&=\left\{x \in \mathbb{R}^6 \mid\left\|p_1 - p_2 \right\| \geq \beta\right\},\\
%    \mathcal{S}_z&= (E(\mathcal{S}_x^c))^c = \left\{z \in \mathbb{R}^3 \mid\left\|\left(z_1, z_2\right)\right\| \geq\beta\right\}.
%\end{align*}
We design a latent CBF $\phi_{\beta'}(z)$ which operates on $z \in \mathbb{R}^3$ to render a subset $\mathcal{C}_z\subseteq\mathcal{S}_z$ forward invariant under $f_z$
via the following latent CBF quadratic program (CBF-QP): %\pi_\beta(z; \theta_2)=\mathrm{arg\,min}_{u\in\mathcal{U}}|u - u_{\text{nom}}|^2$ such that $\langle \nabla \phi_\beta(z), f_z(z, u; \theta_2) \rangle + \alpha \phi_\beta(z) \geq 0$,
% With a valid $\phi_{\beta'}(z)$, we can implement a safety filter by solving the latent CBF quadratic program (CBF-QP):
\begin{align*}
\label{eq:latent_cbf_ineq}
\pi_\beta(z; \theta_2) =\;
& \underset{u\in\mathcal{U}}{\mathrm{arg\,min}} 
& & |u - u_{\text{nom}}|^2   \notag\\
& \mathrm{s.t.}
& &\langle \nabla \phi_\beta(z), f_z(z, u; \theta_2) \rangle + \alpha \phi_\beta(z) \geq 0,
\end{align*}
where we assume $u_\text{nom}$ is generated by a nominal controller that achieves task-specific goals regardless of safety (e.g., proportional control that steers the car towards a goal). 
% We set $\alpha=1$ for the following experiments.

We propose the candidate latent CBF $\phi_{\beta'}(z)=\left\|\left(z_1, z_2\right)\right\|-{\beta'}$ 
over $\mathcal{D}_z = \R^3 \setminus \{0\}$.
% \begin{align*}
%     \phi_{\beta'}(z)=\left\|\left(z_1, z_2\right)\right\|-{\beta'}, \quad \mathcal{D}_z = \{z \in \R^3 \mid (z_1, z_2) \neq 0 \}.
% \end{align*}
% Define the set $\mathcal{D}_z := \R^3 \setminus \{0\}$.
%
We prove in \toggletext{Appendix~\ref{sec:appendix:valid_CBF_proof}}{the full version} that if the input bound
$B_u \geq r \sqrt{3(1+\ell^2)} (1 + \alpha \beta')$,
then for all $z \in \mathcal{D}_z$,
$\sup_{u \in \mathcal{U}} \langle \nabla \phi_{\beta'}(z), f_z(z, u; \theta_2) \rangle + \alpha \phi_{\beta'}(z) \geq 0$.
Hence, the safety set $\mathcal{C}_z = \{ z \in \mathcal{D}_z \mid \phi_{\beta'}(z) \geq 0 \}$ is forward invariant under 
$f_z^{(\pi_{\beta'})}(z; \theta_2) := f_z(z, \pi_{\beta'}(z); \theta_2)$.
Next, it is straightforward to show
that the 
forward conjugacy error (cf.~\Cref{def:continuous_time_dynamics_error}) 
of our setup is bounded by $\gamma = B_d$. Additionally, $\phi_{\beta'}$ is $1$-Lipschitz. Hence, 
if
$\mathcal{C}_z^\gamma := \{ z \in \mathcal{D}_z \mid \phi_{\beta'}(z) \geq -B_d/\alpha \} \subseteq \mathcal{S}_z$, then by 
\Cref{continuous_cbf_proposition} the set
$\mathcal{C}_x^\gamma = \{ x \in \R^6 \mid \| p_1 - p_2 \| \geq \beta' - B_d/\alpha, \,\, x_6 = \theta_2 \}$ is forward invariant under the dynamics $f^{(\pi_{\beta'})}$.
For $\mathcal{C}_z^\gamma \subseteq \mathcal{S}_z$ to occur, we simply take $\beta' \geq \beta + B_d/\alpha$.

We illustrate this example in \Cref{fig:2car_avoidance}. 
Numerically, we choose $\alpha=2$, $B_d=1$, $\beta = 4.5$, $\beta' = 5$, and $B_u=5$. The disturbance is chosen adversarially as $d(x) = - B_d (p_1 - p_2)/\| p_1 - p_2 \|$,
which encourages the relative position between the cars to 
decrease.
The nominal control $u_{\mathrm{nom}}$
is a simple proportional controller to encourage the active car
to move towards the initial position of the passive car;
we arrange the initial condition so that 
using only $u_{\mathrm{nom}}$ for control will result in a safety violation.
We see in \Cref{fig:2car_avoidance} that the behavior predicted
by our theory is precise; by designing $\beta'$ to provide a robust safety margin as required by \Cref{continuous_cbf_proposition}, the active car is able to reach its target goal without exiting the safe set $\mathcal{S}_x$.
}

\toggletext{
% FULL VERSION VVV
\begin{figure}
    \centering
    % full version width: 0.48
    \hspace{-8pt}\includegraphics[width=0.48\textwidth]{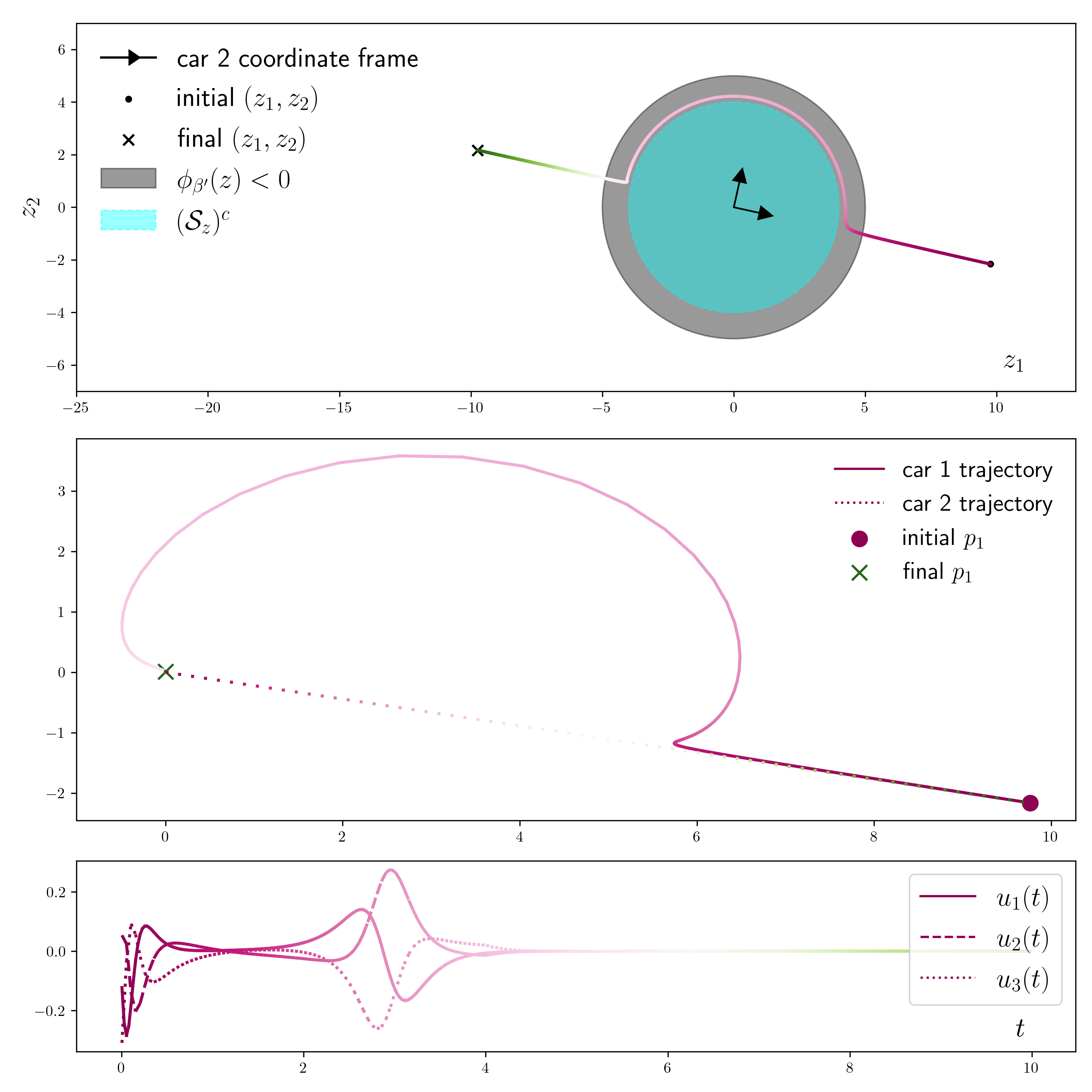}
    \caption{{Numerical simulation of 2-car avoidance system.} Trajectories and plots are color-coded to reflect time. \textit{Top}: Latent space trajectory. The gray circle corresponds to $\{z\mid\phi_{\beta'}(z)<0\}$ and the cyan circle corresponds to $\mathcal{S}_z^c$. \textit{Middle}: State space trajectory resulting from applying the latent-CBF filtered control to the 6D dynamics. \textit{Bottom}: Control signal over time. 
    % \textit{Bottom right}: $\|z(t)\|-(\beta'-2B_d/\alpha)$ over time.} 
    }
    \label{fig:2car_avoidance}
\end{figure}
}{
% CDC VERSION VVV
\begin{figure}
    \centering
    % full version width: 0.48 
    \hspace{-8pt}\includegraphics[width=0.40\textwidth]{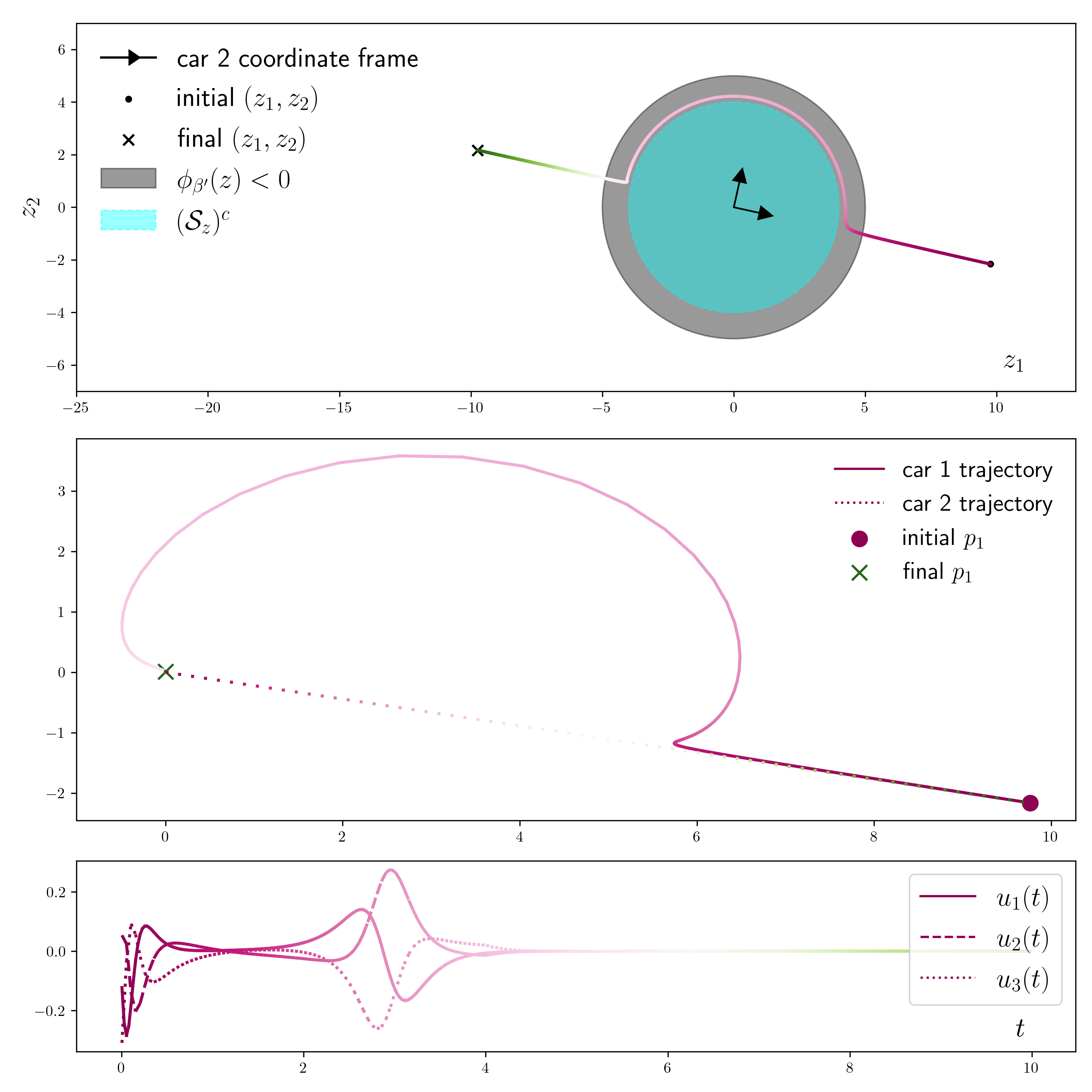}
    \caption{\small{{Numerical simulation of 2-car avoidance system.} Trajectories and plots are color-coded to reflect time. \textit{Top}: Latent space trajectory. The gray circle corresponds to $\{z\mid\phi_{\beta'}(z)<0\}$ and the cyan circle corresponds to $\mathcal{S}_z^c$. \textit{Middle}: State space trajectory resulting from applying the latent-CBF filtered control to the 6D dynamics. \textit{Bottom}: Control signal over time.}
    %\caption{\small{\textcolor{blue}{{2-car avoidance system:} color-gradient reflects time. \textit{Top}: Latent space trajectory. The gray circle corresponds to $\{z\mid\phi_{\beta'}(z)<0\}$, cyan circle corresponds to $\mathcal{S}_z^c$. \textit{Middle}: Trajectory resulting from applying the latent-CBF filtered control to the 6D dynamics. \textit{Bottom}: Control signal. }}
    % \textit{Bottom right}: $\|z(t)\|-(\beta'-2B_d/\alpha)$ over time.} 
    }
    \label{fig:2car_avoidance}
\end{figure}
}
% For the numerical experiment we set $\alpha=2, B_d=1, \beta'=\beta=5$.
\toggletext{
\section{Conclusion and Future Work}
% \textcolor{red}{TODO: write me}
% FULL VERSION VVV
We study the use of control designs based on low-dimensional 
latent representations. Our work identifies approximate conjugacy conditions involving both the encoder/decoder and latent/original dynamics which enable systems analysis. We utilize our definitions to transfer both stability and set-invariance guarantees from the latent system back to the original
system of interest. 

Our work opens up many exciting areas for future study. Some immediate directions include a deeper understanding of the
type of convergence behavior which can be ensured in latent designs;
for example, in our experimental evaluation (specifically
in \Cref{sec:experiments:cartpole}), we see that our latent control law applied to the true system ensures further convergence even after entering the zero-manifold $E^{-1}(0)$, which our theory cannot capture. Another interesting direction involves our empirical observation in \Cref{sec:experiments:cartpole}
that learning latent dynamics using multi-step conjugacy losses is critical to obtaining a low single-step conjugacy error. Thus, extending our theory to multi-step stability criteria using generalized Lyapunov functions (cf.~\cite{furnsinn2025generalized,long2025certifying}) might decrease conservatism and augment the efficacy of our results.

More broadly, extending our theory to settings of partial observability, e.g., control from visual observations, is of immediate relevance especially for modern control stacks which utilize many high-dimensional sensor inputs (multiple RGB cameras, point clouds from LiDAR, etc.).
Furthermore, broadening the applicability of our analysis to 
techniques such as HJ reachability and MPC is of practical relevance as well. 
% \textcolor{blue}{In the cartpole example, we observe that learning latent dynamics using multi-step conjugacy losses is critical to obtaining a low single-step conjugacy error. Thus, extending our theory to multi-step stability criteria \cite{long2025certifying} might decrease conservatism and augment the efficacy of our results.}
%
Finally, defining appropriate \emph{probabilistic} approximate conjugacy conditions which can be used in a systems analysis is also interesting future work, as it would enable the use of efficient uncertainty quantification techniques such as conformal prediction~\cite{vovk2005algorithmic} for estimating dynamics error.
}{
\section{Conclusion and Future Work}
% \textcolor{red}{TODO: write me}
% CDC VERSION VVV
%We study the use of control designs based on low-dimensional 
%latent representations. Our work identifies approximate conjugacy conditions involving both the encoder/decoder and latent/original dynamics which enable systems analysis. We utilize our definitions to transfer both stability and set-invariance guarantees from the latent system back to the original
%system of interest. 

We study control designs based on low-dimensional 
latent representations, transferring
stability and safety guarantees back
to the original system.
Our work opens up many areas for future study. %Some immediate directions include a deeper understanding of the
%type of convergence behavior which can be ensured in latent designs;
In our experimental evaluation, we see that our latent control law applied to the true system ensures further convergence even after entering the zero-manifold $E^{-1}(0)$, which our theory cannot capture. 
%Another interesting direction involves our empirical observation in \Cref{sec:experiments:cartpole}
%that learning latent dynamics using multi-step conjugacy losses is critical to obtaining a low single-step conjugacy error. Thus, extending our theory to multi-step stability criteria using generalized Lyapunov functions (cf.~\cite{furnsinn2025generalized,long2025certifying}) might decrease conservatism and augment the efficacy of our results.
More broadly, extending our theory to settings of partial observability, e.g., control from visual observations, is of immediate relevance, especially for modern control stacks which utilize many high-dimensional sensor inputs (multiple RGB cameras, point clouds from LiDAR, etc.).
%Furthermore, broadening the applicability of our analysis to 
%techniques such as HJ reachability and MPC is of practical relevance as well. 
% \textcolor{blue}{In the cartpole example, we observe that learning latent dynamics using multi-step conjugacy losses is critical to obtaining a low single-step conjugacy error. Thus, extending our theory to multi-step stability criteria \cite{long2025certifying} might decrease conservatism and augment the efficacy of our results.}
%
%Finally, defining appropriate \emph{probabilistic} approximate conjugacy conditions which can be used in a systems analysis is also interesting future work, as it would enable the use of efficient uncertainty quantification techniques such as conformal prediction~\cite{vovk2005algorithmic} for estimating dynamics error.
}

\bibliographystyle{ieeetr}
\bibliography{ref.bib}

\toggle{
\appendices
\input{l4dc2025_latex_template/appendix}
}

\end{document}

%% file: l4dc2025_latex_template/appendix.tex
\section{Estimating Invariant Subsets of $\mathcal{D}_z$ and $\mathcal{D}_x$}
\label{Dx_and_Dz_details}

Here, we discuss how we 
estimate bounded and (approximately) invariant subsets of $\mathcal{D}_z$ and $\mathcal{D}_x$.
We first compute a bounded subset of $\mathcal{D}_z$ by
gridding an initial set of \emph{latent} initial conditions $z_0$ over a latent hypercube $\mathcal{H}_{r_1}^\ell = \{ z \in \mathbb{R}^{n_z} \mid \| z \|_\infty \leq r_1 \}$ for some constant $r_1>0$, 
and then set $\mathcal{D}_z = \mathrm{conv}\left(\cup_{t=0}^{T} (f^{(\pi)}_z \circ \overset{t}\hdots \circ f^{(\pi)}_z)(\mathcal{H}_{r_1}^\ell)\right)$ 
as the convex-hull of the union of all $t \leq T$ step latent dynamics trajectories, where $T = 300$. In the cartpole case study, $r_1=1.5$. The reasoning here is that the set inside the convex-hull approximates an invariant subset of $\mathcal{D}_z$, whereas the convex-hull, while introducing an over-approximation, allows for efficient checking of set membership.
With this approximate $\mathcal{D}_z$ in place,
we now describe the computation of an approximately forward invariant subset of $\mathcal{D}_x$.
We apply a similar procedure as described for $\mathcal{D}_z$, by first computing a set of initial conditions $\mathcal{X}_0 = E^{-1}(\mathcal{D}_z) \cap \mathcal{H}_{r_2}$, where $\mathcal{H}_{r_2} = \{ x \in \mathbb{R}^{n_x} \mid \| x \|_\infty \leq  r_2\}$ is a hypercube in the original state space.
In the cartpole case study, $r_2=0.5$. Note the computation of $\mathcal{X}_0$ is implemented by gridding over $\mathcal{H}_{r_2}$ and, for each grid point $x$, checking if $E(x) \in \mathcal{D}_z$ (hence the importance of fast set membership calculation). Given the initial conditions $\mathcal{X}_0$, we then set $\mathcal{D}_x = \cup_{t=0}^{T'} (f^{(\pi)} \circ \overset{t} \hdots \circ f^{(\pi)})(\mathcal{X}_0)$. In practice, when $\mathcal{X}_0$ is a finite set of initial conditions, we may compute 
$\mathcal{D}_x = \cup_{t=0}^{T'} ((f^{(\pi)} \circ \overset{t} \hdots \circ f^{(\pi)})(\mathcal{X}_0) \oplus \mathcal{B}_\epsilon)$, for some constant $\epsilon>0$,
where $\oplus$ denotes the Minkowski sum and 
$\mathcal{B}_\epsilon$ denotes an $\ell_\infty$ ball of radius $\epsilon$ around the origin. Because we would like to sample from $\mathcal{D}_x$, the purpose of performing a Minkowski sum with $\epsilon$-balls is to obtain a $\mathcal{D}_x$ with nonzero measure even when $\mathcal{X}_0$ is finite.
We then post-hoc numerically check that the set $\mathcal{D}_x$ as constructed is approximately forward invariant by randomly sampling $x_0 \in \mathcal{D}_x$ and checking that $\cup_{t=0}^{T'}((f^{(\pi)} \circ \overset{t} \hdots \circ f^{(\pi)})(x_0) \in \mathcal{D}_x$.

\section{Proof of Valid CBF for Omni-directional Vehicle}
\label{sec:appendix:valid_CBF_proof}

First, we note that the dynamics matrix for the omni-directional
vehicle are:
\begin{align*}
    G(\theta_1)&= \left[\begin{array}{ccc}
    \cos (\theta_1) & -\sin(\theta_1) & 0 \\
    \sin (\theta_1) & \cos(\theta_1) & 0 \\
    0 & 0 & 1
    \end{array}\right],
    \\
    B&=\left[\begin{array}{ccc}
    0 & r \cos (\pi / 6) & -r \cos (\pi / 6) \\
    -r & r \sin (\pi / 6) & r \sin (\pi / 6) \\
    \ell r & \ell r & \ell r
    \end{array}\right].
\end{align*}

We first prove that the quantity 
$\sigma_{\min}(B^{-1} G(z_3)^T)$ is bounded away from zero uniformly for all $z_3$.
Let 
$$
    R(\theta) = \begin{bmatrix} \cos(\theta) & -\sin(\theta) \\ \sin(\theta) & \cos(\theta) \end{bmatrix}
$$
denote the planar rotation matrix for $\theta$.
We observe:
\begin{align*}
    &\sigma^2_{\min}(B^{-1} G(z_3)^T) \\
    &= \lambda_{\min}( G(z_3) B^{-T} B^{-1} G(z_3)^T) \\
    &\geq \sigma_{\min}^2(B^{-1}) \lambda_{\min}(G(z_3) G(z_3)^T) \\
    &= \frac{1}{\sigma_{\max}^2(B)} \lambda_{\min}\left( \begin{bmatrix} R(z_3) & 0 \\ 0 & 1 \end{bmatrix}\begin{bmatrix} R(z_3)^T & 0 \\ 0 & 1 \end{bmatrix}  \right) \\
    &= \frac{1}{\sigma_{\max}^2(B)} \lambda_{\min}\left( \begin{bmatrix} R(z_3)R(z_3)^T & 0 \\ 0 & 1 \end{bmatrix} \right) \\
    &= \frac{1}{\sigma^2_{\max}(B)},
\end{align*}
since $R(z)R(z)^T = I_2$ for all $z$.
From this, we conclude that $\sigma_{\min}(B^{-1} G(z_3)^T) \geq 1/\sigma_{\max}(B)$.
Now we can prove that the CBF inequality holds if we set the input bound large enough.
Recall that $\mathcal{U} = \{ u \in \R^3 \mid \| u \|_\infty \leq B_u \}$.
We observe that, for any $z \in \R^3$ with 
$z_{12} := (z_1, z_2) \neq 0$ and $\bar{z}_{12} := z_{12}/\| z_{12} \|$, putting $M(z) := G(z_3) B^{-T}$ and
$v_2 := (\cos(\theta_2), \sin(\theta_2))$,
\begin{align*}
    & \sup_{ u \in \mathcal{U} } \langle \nabla \phi_{\beta'}(z), f_z(z, u; \theta_2) \rangle + \alpha \phi_{\beta'}(z) \\
    &= \sup_{ u \in \mathcal{U} } \langle (\bar{z}_{12}, 0), M(z_3) u \rangle - \langle \bar{z}_{12}, v_2 \rangle + \alpha( \| z_{12} \| - \beta' ) \\
    &= B_u \| M(z_3)^T (\bar{z}_{12}, 0) \|_1 - \langle \bar{z}_{12}, v_2 \rangle + \alpha( \| z_{12} \| - \beta' ) \\
    &\geq B_u \| M(z_3)^T (\bar{z}_{12}, 0) \| - \langle \bar{z}_{12}, v_2 \rangle + \alpha( \| z_{12} \| - \beta' ) \\
    &\geq B_u \sigma_{\min}( M(z_3)^T ) \| (\bar{z}_{12}, 0) \| - \| \bar{z}_{12} \| \|v_2 \| - \alpha \beta' \\
    &= B_u \sigma_{\min}( B^{-1} G(z_3)^T ) - 1 - \alpha\beta' \\
    &\geq B_u / \sigma_{\max}(B)  - 1 - \alpha\beta'.
\end{align*}
Hence, if 
$$
    B_u \geq \sigma_{\max}(B) (1+ \alpha \beta'),
$$
then we have for all $z \in \R^3$ with $z_{12} \neq 0$,
$$
    \sup_{ u \in \mathcal{U} } \langle \nabla \phi_{\beta'}(z), f_z(z, u; \theta_2) \rangle + \alpha \phi_{\beta'}(z) \geq 0.
$$
Note that a simple bound on $\sigma_{\max}(B)$ is:
$$
    \sigma_{\max}(B) \leq \| B \|_F = r \sqrt{3(1+\ell^2)},
$$
from which we derive the following sufficient condition:
$$
    B_u \geq r \sqrt{3(1+\ell^2)}(1 + \alpha \beta').
$$
%\color{blue}
\section{Forward Completeness of Omni-directional Vehicle}\label{app:omni_forward_completeness}
\begin{prp}[Forward Completeness of Omni-Directional Vehicle System under Bounded Inputs]\label{prp:omni_forward_completeness}
The omni-directional vehicle system described in \Cref{sec:omni-directional_vehicle} is forward complete for bounded inputs $\| u \| \leq B_u$ and bounded disturbances $\|d(x)\|\leq B_d$.
\end{prp}
\noindent\emph{Proof.} We focus our attention on the active car. Let the second subscript of state denote time. The passive car moves in a straight line with constant velocity, and so is trivially forward complete, satisfying $\|p_{2,t}\|\leq\|p_{2,0}\|+t$ and $|\theta_{2,t}|=|\theta_{2,0}|$.\par
Recall that the active car has dynamics
\begin{gather}
\dot{q}_1=G(\theta_1)B^{-T}u+d(x).
\end{gather}
% where $G(\theta_1)$ is invertible for all $\theta_1$, and $B$ is invertible. 
We first observe that $G(\theta_1)^T G(\theta_1) = I_3$.
Hence we have:
\begin{align*}
    \sigma_{\max}(G(\theta_1) B^{-T}) &= \sqrt{\lambda_{\max}( B^{-1} G(\theta_1)^T G(\theta_1) B^{-T})} \\
    &= \sigma_{\max}(B^{-T}).
\end{align*}
Therefore:
\begin{align*}
    \| \dot{q}_1 \| &\leq \| G(\theta_1) B^{-T} u \| + \| d(x) \| \\
    &\leq \sigma_{\max}( G(\theta_1) B^{-T} ) \| u \| + \| d(x) \| \\
    &\leq \sigma_{\max}( B^{-T} ) B_u + B_d.
\end{align*}
Thus by integrating $q_{1,t}$ (recall, second subscript is time):
\begin{align*}
    \| q_{1,t} \| &= \left\| q_{1,0} + \int_0^t \dot{q}_{1,s} \mathrm{d} s \right\| \\
    &\leq \| q_{1,0} \| + \int_0^t \| \dot{q}_{1,s} \| \mathrm{d} s \\
    &\leq \| q_{1,0} \| + \left[ \sigma_{\max}( B^{-T} ) B_u + B_d \right] t.
\end{align*}
Therefore, the system is is forward complete.$\quad\blacksquare$
%
% %
% % Consider $\|\dot q_1\|$, 
% % \begin{gather}
% %     |\dot q_1^{(i)}|\leq \|\dot q_1\|\leq\|G(\theta_1)B^{-T}u\|+\|d(x)\|\\
% %     \leq\left(3\sigma_\mathrm{max}(G(\theta_1)B^{-T})\right)^{\frac{1}{2}}B_u +B_d\ , 
% % \end{gather}
% % where $0\leq\sigma_\mathrm{max}(G(\theta_1)B^{-T})$ is the maximum singular value of $G(\theta_1)B^{-T}$. 
% By the triangle inequality, we have
% \begin{gather}
% \|q_1(t)\|=\left\|\int\dot q_1 dt\right\|\leq\int\left\|\dot q_1\right\|dt\leq\\
% \leq \|q_1(0)\|+\left[\sqrt{3\sigma_\mathrm{max}(G(\theta_1)B^{-T})}B_u+B_d\right]t
% \end{gather}
% Thus, the components $q_1^{(i)}$ of the active car satisfy $|q_1^{(i)}(t)|\leq \|q_1(0)\|+\big[\sqrt{3\sigma_\mathrm{max}(G(\theta_1)B^{-T})}B_u+B_d\big]t$ and the system is forward complete.$\quad\blacksquare$

\section{Forward Completeness of Cartpole}\label{app:cartpole_forward_completeness}
We first derive a useful lemma that relates power to control input and velocity. We then use the lemma to show that a common class of force-actuated mechanical systems are forward complete. 
Finally, we use these results to prove that the cartpole system
described by \eqref{eq:cartpole} is forward complete.
% Finally, we prove that the cartpole system described by \cref{eq:cartpole} is one of these systems.

\begin{lmm}[Hamiltonian-Power Relation]\label{lmm:hamiltonian-power}
Consider a locally Lipschitz system with generalized positions %$q(t) \in \R^n$ 
$q_t\in\R^n$ and velocities %$\dot q(t) \in \R^n$
$\dot q_t\in\R^n$, time-independent kinetic and potential energy functions $T(q,\dot q)=\frac{1}{2}\langle\dot q, M(q)\dot q\rangle$ and $V(q)$, and total energy $E(q,\dot q)=T(q,\dot q)+V(q)$, where $M(q)$ is a $d \times d$ symmetric positive definite matrix. If the components $q$ of the system are actuated with forces $u_t$, then we can express the power $\dot E_t$ of the system as 
\begin{equation}
    %\dot E(t)=\sum_{i=1}^{n} u_i(t)\dot q_i(t).
    \dot E_t=\langle u_t,\dot q_t\rangle
\end{equation}
\end{lmm}
\noindent\emph{Proof.} %Define the Hamiltonian to be $H(p,q)=T(q,\dot{q})+V(q)$, i.e. the total energy. We assume the Hamiltonian (and subsequent Lagrangian to be time-independent). 
We first recall the classic definitions of Lagrangian and Hamiltonian:
\begin{align*}
    L(q_t, \dot{q}_t) &= T(q_t, \dot{q}_t) - V(q_t), \\
    %H(p, q) &= \sum_i p_i\dot{q}_i-L(q,\dot{q}), \quad p_i = \frac{\partial L}{\partial \dot q_i}.
    H(p_t, q_t) &= \langle p_t,\dot q_t\rangle-L(q_t,\dot{q}_t), \quad p_t = \nabla_{\dot q} L(q_t,\dot q_t).
\end{align*}
The total derivatives of both the Lagrangian and Hamiltonian with respect to time are given by:
\begin{align*}
    %\dot L(q,\dot{q}) &= \sum_i\left(\frac{\partial L}{\partial \dot q_i}\ddot q_i+\frac{\partial L}{\partial q_i}\dot q_i\right) = \sum_i\left(p_i\ddot q_i+\frac{\partial L}{\partial q_i}\dot q_i\right), \\
    \dot L(q_t,\dot{q}_t) &=\langle\nabla_{\dot q}L, \ddot q_t\rangle+\langle\nabla_q L,\dot q_t\rangle = \langle p_t,\ddot q_t\rangle+\langle\nabla_q L,\dot q_t\rangle, \\
    \dot H(p_t,q_t) &= \langle\dot p_t,\dot q_t\rangle+\langle p_t,\ddot q_t\rangle-\dot L(q_t,\dot q_t) = \langle\dot p_t-\nabla_q L,\dot q_t\rangle, 
\end{align*}
where $p_t$ are generalized momenta. Because $p_t=\nabla_{\dot q} L$, we have $\dot p_t=\frac{\mathrm{d}}{\mathrm{d} t}\left[\nabla_{\dot q} L\right]$ and therefore 
\begin{align*}
     \dot H(p_t,q_t)=\left\langle\frac{\mathrm{d}}{\mathrm{d} t}\left[\nabla_{\dot q} L\right]-\nabla_q L,\dot q_t\right\rangle.
\end{align*}
Observe that the left argument of the above inner product is simply the Euler-Lagrange equation\cite{marsden2013introduction} for $q$, which equals $u_t$ when the components $q$ are force-actuated, hence:
\begin{align*}
    \dot H(p_t, q_t) = \langle u_t ,\dot{q}_t\rangle.
\end{align*}
To conclude, we will show that the Hamiltonian is equal to the energy, i.e. $H(p_t, q_t) = T(q_t, \dot{q}_t) + V(q_t) = E(q_t, \dot{q}_t)$.
Indeed, observe that:
\begin{align*}
    H(p_t, q_t) &= \langle \nabla_{\dot{q}} L(q_t, \dot{q}_t), \dot{q}_t \rangle - L(q_t, \dot{q}_t) \\
    &\stackrel{(a)}{=} \langle \nabla_{\dot{q}} T(q_t, \dot{q}_t), \dot{q}_t \rangle - T(q_t, \dot{q}_t) + V(q_t) \\
    &= \langle M(q) \dot{q}, \dot{q} \rangle - \frac{1}{2} \langle\dot{q}_t,M(q_t)\dot{q}_t\rangle + V(q_t) \\
    &= \frac{1}{2}\langle\dot{q}_t,M(q_t)\dot{q}_t\rangle + V(q_t) = T(q_t, \dot{q}_t) + V(q_t),
\end{align*}
where (a) holds since $V(q_t)$ is not a function of $\dot{q}$.
%
% Because $T(q,\dot q)=\frac{1}{2}\dot q^T M(q)\dot q$ for symmetric positive definite $M(q)$, and $V(q)$ does not depend on $\dot q$ or $t$, the Hamiltonian is the total system energy, i.e., $H(p,q)=T(q,\dot q)+V(q)=E(q,\dot q)$. Therefore, 
% \begin{align}
%     \dot E(t)=\sum_i u_i(t)\dot q_i(t),
% \end{align}
% which completes the proof.
We have thus shown that $\dot{E}(q_t, \dot{q}_t) = \langle u_t,\dot{q}_t\rangle$,
which completes the proof. $\quad\blacksquare$

% Since the generalized momenta $p_i$ satisfy $p_i=\frac{\partial L}{\partial \dot q_i}$, we have:
% \begin{align}
%     \dot L(q,\dot q)=\sum_i\left(p_i\ddot q_i+\frac{\partial L}{\partial q_i}\dot q_i\right) \\
%     \implies \dot H(p,q)=\sum_i\left(\dot p_i-\frac{\partial L}{\partial q_i}\right)\dot q_i
% \end{align}

% By definition, the Hamiltonian satisfies
% \begin{align}
%     H(p,q) &= \sum_i p_i\dot{q}_i-L(q,\dot{q}), \\
%     \dot H(p,q) &= \sum_i(\dot p_i\dot q_i+p_i\ddot q_i)-\dot L(q,\dot q).
% \end{align}
% The total derivative of the Lagrangian with respect to time is 
% \begin{align}
%     \dot L(q,\dot{q})=\sum_i\left(\frac{\partial L}{\partial \dot q_i}\ddot q_i+\frac{\partial L}{\partial q_i}\dot q_i\right)
% \end{align}
% Note that the generalized momenta $p_i$ satisfy $p_i=\frac{\partial L}{\partial \dot q_i}$, so we can write 
% \begin{align}
%     \dot L(q,\dot q)=\sum_i\left(p_i\ddot q_i+\frac{\partial L}{\partial q_i}\dot q_i\right)\\
%     \implies \dot H(p,q)=\sum_i\left(\dot p_i-\frac{\partial L}{\partial q_i}\right)\dot q_i
% \end{align}

\begin{prp}[Forward-Completeness]\label{prp:forward_completeness}
Consider a locally Lipschitz, second-order system $\ddot{q}=f(q, \dot q, u)\in\mathbb{R}^n$ with kinetic energy given by $T(q,\dot q)=\frac{1}{2}\langle\dot q,M(q)\dot q\rangle$ for a uniformly\footnote{That is, there exists a $\mu > 0$ such that $\lambda_{\min}(M(q)) \geq \mu$ for all $q$.}  symmetric positive definite $M(q)$ and non-negative potential energy $V(q)$. 
% If the control input $u(t)$ satisfies $-u_\mathrm{max}\leq u_i(t)\leq u_\mathrm{max}$ for some $u_\mathrm{max}>0$, all $i$, and all $t\geq 0$, 
If the input signal $u_t$ satisfies $\| u_t \| \leq B_u$ for all $t \geq 0$, 
then the trajectory $(q_t, \dot{q}_t)$ is forward complete.
% and system trajectories exist for all time (no finite-time blowup can occur).
\end{prp}
\noindent\emph{Proof.} Since the system satisfies the requirements of \Cref{lmm:hamiltonian-power}, we have that:
\begin{align*}
    \dot E_t=\langle u_t,\dot q_t\rangle.
\end{align*}
Since potential energy is assumed to be non-negative,
\begin{align*}
    E_t\geq\frac{1}{2}\langle\dot{q}_t, M(q_t) \dot{q}_t\rangle\geq \frac{\mu}{2} \| \dot{q}_t \|^2 ,
\end{align*}
where the uniformly PSD assumption on $M(q)$ implies existence of a positive $\mu$
which is independent of $q$.
This inequality implies that 
$\| \dot{q}_t \| \leq \sqrt{ 2 E_t / \mu }$.
%
%
% As $M(q)$ is symmetric and positive definite for all $q$, by the Rayleigh-Ritz theorem we have that 
% \begin{gather}
%     0<\lambda_\mathrm{min}\leq\frac{\dot q^TM(q)\dot q}{\dot q^T\dot q}\leq\lambda_\mathrm{max}\\
%     \lambda_\mathrm{min}\|\dot q\|^2\leq \dot q^T M(q)\dot q\leq 2E(t)\\
%     \|\dot q\|\leq \sqrt{\frac{2E(t)}{\lambda_\mathrm{min}}}
% \end{gather}
Applying the uniform bound on $\| u_t \|$, we have by Cauchy-Schwarz,
\begin{align*}
    \dot{E}_t &= \langle u_t, \dot{q}_t \rangle \leq \| u_t \| \| \dot{q}_t \| \leq B_u \| \dot{q}_t \| \\
    &\leq B_u \sqrt{2 E_t / \mu}.
\end{align*}
% \begin{gather}
%     \dot E(t)\leq u_\mathrm{max}\sum_i\dot q_i(t)\leq u_\mathrm{max}\sqrt{n}\|\dot q\|\ .
% \end{gather}
Hence the following differential inequality holds:
\begin{align*}
    \frac{\mathrm{d}}{\mathrm{d} t} \sqrt{E_t} = \frac{\dot{E}_t}{2\sqrt{E_t}} \leq \frac{B_u}{\sqrt{2 \mu}}.
\end{align*}
Integrating this bound yields:
\begin{align*}
    \sqrt{E_t} = \sqrt{E_0} + \int_0^t \frac{\mathrm{d}}{\mathrm{d}s} \sqrt{E_s} \mathrm{d} s \leq \sqrt{E_0} + \frac{B_u t}{\sqrt{2\mu}}.
\end{align*}
Thus, we have shown that:
\begin{align*}
    \| \dot{q}_t \| \leq \sqrt{\frac{2}{\mu}}\left[ \sqrt{E_0} + \frac{B_u t}{\sqrt{2\mu}} \right] =: A + B t,
\end{align*}
Now integrating this bound on the velocity,
\begin{align*}
    \| q_t \| &= \left\| q_0 + \int_0^t \dot{q}_s \mathrm{d} s \right\| \\
    &\leq \| q_0 \| + \int_0^t \| \dot{q}_s \| \mathrm{d} s \\
    &\leq \| q_0 \| + \int_0^t (A + Bs) \mathrm{d} s \\
    &= \|q_0\| + A t + \frac{Bt^2}{2}.
\end{align*}
%
% Consider the ODE
% \begin{gather}
%     \frac{d}{dt}\sqrt{E(t)}=\frac{\dot{E}}{2\sqrt{E}}\leq \frac{u_\mathrm{max}\sqrt{n}\|\dot q\|}{2\sqrt{E}}\\
%     \leq u_\mathrm{max}\sqrt{\frac{n}{2\lambda_\mathrm{min}}}
% \end{gather}
% Integrating yields
% \begin{gather}
%     \sqrt{E(t)}\leq \sqrt{E(0)}+u_\mathrm{max}\sqrt{\frac{n}{2\lambda_\mathrm{min}}}t\\
%     E(t)\leq E(0)+u_\mathrm{max}\sqrt{\frac{2n}{\lambda_\mathrm{min}}}E(0)t+\frac{nu_\mathrm{max}^2}{2\lambda_\mathrm{min}}t^2\ ,
% \end{gather}
% i.e. $E(t)\geq0$ is upper-bounded by a quadratic function in time. Because we established $\|\dot q\|\leq\mathcal{O}\big(E^{\frac{1}{2}}(t)\big)$, we also have $\|\dot q\|\leq\mathcal{O}(t)$:
% \begin{gather}
%     |\dot q_i| \leq \|\dot q \| \leq\sqrt{\frac{2E(t)}{\lambda_\mathrm{min}}}\\
%     |\dot q_i|\leq \sqrt{\frac{2}{\lambda_\mathrm{min}}}\left[\sqrt{E(0)}+u_\mathrm{max}\sqrt{\frac{n}{2\lambda_\mathrm{min}}}t\right]
% \end{gather}
Therefore,
$q_t$ grows at most quadratically in $t$
and $\dot{q}_t$ grows at most linearly in $t$,
which completes the proof. $\quad\blacksquare$

\begin{prp}[Forward Completeness of Cartpole under Bounded Inputs]\label{prp:cartpole_forward_completeness}The system described by \Cref{eq:cartpole} is forward complete when $| u_t | \leq u_\mathrm{max}$  and $m_c, m_p, l>0$.
\end{prp}
\noindent\emph{Proof.} The cartpole system is continuously differentiable and therefore locally Lipschitz continuous. We first derive the total energy of the cartpole system described by \Cref{eq:cartpole}, where the pole is a rod of length $2l$ with uniform density. The total energy is the sum of kinetic and potential components, i.e., 
$$E_t=T^{(c)}_t+T^{(p)}_t+V^{(p)}_t,$$ 
where $T^{(c)}, T^{(p)}$ are kinetic energies of cart and pole respectively, and $V^{(p)}$ is the potential energy of the pole:
\begin{align*}
    T^{(c)}_t&=\frac{1}{2}m_c\dot{x}_t^2, \\
    T^{(p)}_t&=T_t^{(\mathrm{p, trans.})}+T_t^{(p,\mathrm{rot.})}\\
    &=\frac{m_p}{2}\left(\frac{\mathrm{d}}{\mathrm{d}t}\bigg[x_t+l\sin\theta_t\bigg]^2+\frac{\mathrm{d}}{\mathrm{d}t}\bigg[l\cos\theta_t\bigg]^2\right)+\frac{I_\mathrm{cm}}{2}\dot{\theta}_t^2\\
    &=m_p\left(\frac{\dot{x}_t^2}{2}+l\dot x_t\dot\theta_t\cos\theta_t+\frac{l^2}{2}\dot\theta_t^2\right)+\frac{m_pl^2}{12}\dot\theta_t^2,
\end{align*}
where $I_\mathrm{cm}=(1/12)m_pl^2$ is the moment of inertia about the center of mass of the pole. We next define the potential energy of the pole so that it is a non-negative function.
\begin{align*}
    V^{(p)}_t=m_pgl(\cos{\theta_t}+1).
\end{align*}
The total energy $E_t$ is therefore:
\begin{align*}
    E_t &= \frac{(m_c+m_p)}{2}\dot{x}_t^2+m_p\left(l\dot x_t\dot\theta_t\cos\theta_t+\frac{l^2}{2}\dot\theta_t^2+\frac{l^2}{12}\dot\theta_t^2\right)\\
    &\qquad+m_pgl(\cos\theta_t+1).
\end{align*}
Observe that the potential energy $V^{(p)}_t$ is non-negative and time-independent. Furthermore, the kinetic energy is time-independent and can be expressed as the following quadratic form:
\begin{align*}
T_t&=T^{(c)}_t+T^{(p)}_t=\begin{bmatrix}
    \dot x_t\\
    \dot \theta_t
\end{bmatrix}^T
\begin{bmatrix}
    \frac{(m_c+m_p)}{2} & \frac{m_pl\cos\theta_t}{2} \\
    \frac{m_pl\cos\theta_t}{2} & \frac{7m_pl^2}{12}
\end{bmatrix}
\begin{bmatrix}
    \dot x_t\\
    \dot \theta_t
\end{bmatrix}\\
&=: \langle\dot q_t,M(q_t)\dot q_t\rangle.
\end{align*}
To finish the proof, we need to show that $M(q)$ is uniformly positive definite for all $q$.
Let us define the family of matrices indexed by scalar $\rho \in [-1, 1]$:
$$
    M_\rho := \begin{bmatrix}
    \frac{(m_c+m_p)}{2} & \frac{m_pl\rho}{2} \\
    \frac{m_pl\rho}{2} & \frac{7m_pl^2}{12}
    \end{bmatrix},
$$
so that $M(q) = M_{\cos(\theta)}$.
We have that:
\begin{align*}
    \mathrm{tr}(M_\rho) &= \frac{m_c+m_p}{2}+\frac{7m_pl^2}{12}>0, \\
    \mathrm{det}(M_\rho) &= \frac{7}{24}\left[m_p^2l^2-\frac{m_p^2l^2\rho^2}{4}+m_pm_cl^2\right]>0,
\end{align*}
Since both $\mathrm{tr}(M_\rho)$ and $\mathrm{det}(M_\rho)$ are positive,
this shows that $M_\rho$ is positive definite for a $\rho \in [-1, 1]$.
Furthermore, since the function $\rho \mapsto \lambda_{\min}(M_\rho)$ is continuous, we have that $\inf_{\rho \in [-1, 1]} \lambda_{\min}(M_\rho)$
is achieved by some $\rho_\star \in [-1, 1]$, and hence 
$\inf_{\rho \in [-1, 1]} \lambda_{\min}(M_\rho) = \lambda_{\min}(M_{\rho_\star}) > 0$, showing that $M(q)$ is uniformly positive definite.
%
% Note that the trace of $M(q)$ is given by $\mathrm{tr}(M(q))=\frac{m_c+m_p}{2}+\frac{7m_pl^2}{12}>0$, and that the determinant is $\mathrm{det}(M(q))=\frac{7}{24}\left[m_p^2l^2-\frac{m_p^2l^2\cos^2\theta}{4}+m_pm_cl^2\right]>0$. Because both the determinant and trace are positive, the matrix $M(q)$ is positive definite.

We have thereby established that the cartpole system satisfies the assumptions of \Cref{lmm:hamiltonian-power}: time-independent Hamiltonian; and also the assumptions of \Cref{prp:forward_completeness}: uniformly symmetric positive-definite $M(q)$, non-negative $V(q)$. As a consequence of \Cref{prp:forward_completeness}, the cartpole system described in \eqref{eq:cartpole} is forward complete.$\quad\blacksquare$